\documentstyle{elsart}     
\begin{document}

\renewcommand{\theequation}{\thesection.\arabic{equation}}

\def\e{\mbox{e}}
\def\ib{\,\mbox{i}\,}
\def\is{\,\mbox{\scriptsize i}\,}
\def\la{\lambda}
\def\th{\hbox{th}}
\def\te{\vartheta_1}
\def\td{\vartheta_3}
\def\tv{\vartheta_4}
\def\s{\mbox{sh}}
\def\c{\mbox{ch}}
\def\case#1#2{{\textstyle{#1\over #2}}}
\def\W#1#2#3#4#5{W #1 \! \left(\hspace{-1mm}
         \begin{array}{cc}#5 & #4 \\ #2 & #3 \end{array}
         \hspace{-1mm}\right)}
\def\F{{\cal F}}
\def\G{{\cal G}}

\begin{frontmatter}

\title{Correlation lengths and $E_8$ mass spectrum\\ of the 
dilute $A_3$ lattice model\thanksref{talk}}
\thanks[talk]{Expanded version of a talk presented at the
International Workshop on Statistical Mechanics and Integrable
Systems (Coolangatta, Australia, July 1997).} 

\author[Canberra]{M.T. Batchelor} and
\author[LaTrobe]{K.A. Seaton}
\address[Canberra]{Department of Mathematics,
School of Mathematical Sciences,
Australian National University, Canberra ACT 0200, Australia}
\address[LaTrobe]{School of Mathematics, La Trobe University,
                 Bundoora, Victoria 3083, Australia}

\begin{abstract}
The exact perturbation approach is used to derive the 
elementary correlation lengths $\xi_i$ and related mass gaps $m_i$ of 
the two-dimensional dilute $A_L$ lattice model in regimes 1 and 2 for 
$L$ odd from the Bethe Ansatz solution. 
In regime 2 the $A_3$ model is the $E_8$ lattice realisation of the
two-dimensional Ising model in a magnetic field at $T=T_c$.
The calculations for the $A_3$ model in regime 2 start from
the eight thermodynamically significant string types found
in previous numerical studies. These string types are seen to be
consistent in the ordered high field limit. The eight masses
obtained reduce with the approach to criticality to the $E_8$ 
masses predicted by Zamolodchikov,
thus providing a further direct lattice determination of the 
$E_8$ mass spectrum.
\end{abstract}

\end{frontmatter}

\section{Introduction}

\setcounter{equation}{0}

There is an intimate relationship between conformal field theory, 
integrable field theory and integrable lattice models in
statistical mechanics \cite{CH}. In particular, massive integrable 
field theory
can be considered as conformal field theory perturbed by some
scalar relevant operator. The perturbed Hamiltonian is
\begin{equation}
H = H_c + \lambda \int \Phi(r) d^2(r) .
\end{equation}
The canonical example is the Ising model. The
$\Phi_{(1,3)}$ perturbation is thermal and introduces a single
correlation length into the system. The off-critical system is
described by scalar combinations of massive Majorana fermions
of mass proportional to the inverse of the correlaton length.
Here the integrable field theory can be considered as $c=\half$
conformal field theory perturbed by the energy density $\Phi_{(1,3)}$
having the conformal dimensions $(\half,\half)$. 

In a remarkable advance,
Zamolodchikov \cite{Za,Zb} considered the $\Phi_{(1,2)}$ magnetic
perturbation and showed that there are a number of nontrivial
local integrals of motion and thus an integrable field theory. 
In this case the $c=\half$ conformal field theory is perturbed by
the spin operator of dimension $(\case{1}{16},\case{1}{16})$.
Zamolodchikov then conjectured the $S$-matrix and mass spectrum of 
this field theory. The masses coincide with the components
of the Perron-Frobenius vector of the Cartan matrix of the Lie algebra
$E_8$. They are
\begin{equation}
\begin{array}{ll}
m_2/m_1 = 2 \cos \frac{\pi}{5}  & = 1.618~033\ldots \\
m_3/m_1 = 2 \cos \frac{\pi}{30} & = 1.989~043\ldots \\ 
m_4/m_1 = 4 \cos \frac{\pi}{5} \cos \frac{7\pi}{30} & = 2.404~867\ldots \\ 
m_5/m_1 = 4 \cos \frac{\pi}{5} \cos \frac{2\pi}{15} & = 2.956~295\ldots \\ 
m_6/m_1 = 4 \cos \frac{\pi}{5} \cos \frac{\pi}{30}  & = 3.218~340\ldots \\ 
m_7/m_1 = 8 \cos^2 \frac{\pi}{5} \cos \frac{7\pi}{30} & = 3.891~156\ldots \\ 
m_8/m_1 = 8 \cos^2 \frac{\pi}{5} \cos \frac{2\pi}{15} & = 4.783~386\ldots 
\end{array}
\label{masses}
\end{equation}
Very soon after, numerical tests were performed to check that these masses
were indeed present in the lattice model, namely  
the two-dimensional Ising model in a magnetic field at $T=T_c$.
The first few masses were convincingly observed in the related quantum Ising
chain in a magnetic field via finite-size diagonalisation \cite{HS,SZ,H} and
the truncated fermionic state space method \cite{YZ}. 

Zamolodchikov also pointed out the possibility of an integrable off-critical
lattice model corresponding to the integrable magnetic perturbation of the 
field theory.
In a further development, 
an integrable lattice realisation of the $E_8$ Ising model
was provided by the dilute $A_3$ model \cite{WNS}.
In this model the elliptic nome plays the role of magnetic field.
The calculation of the bulk free energy of the dilute $A_3$ model in the
appropriate regime gives the magnetic Ising exponent
$\delta=15$ \cite{WNS}, which also follows from the calculation
of the local height probability \cite{WPSN}. 
The expected Ising magnetic surface exponent $\delta_s = -\case{15}{7}$ 
follows from the excess surface free energy \cite{BFZ}.
The study of the thermodynamics of the dilute $A_3$ model revealed the
entire $E_8$ mass spectrum in the scaling limit \cite{BNW}. In
particular, the resulting integral equations in the thermodynamic Bethe 
Ansatz calculations are those discussed earlier based on the Lie 
algebra $E_8$ \cite{BR}.
The $E_8$ structure in the dilute $A_3$ model in regime 2 has also been 
established by expressing the one-dimensional configuration sums appearing
in the local height probability \cite{WPSN} in terms of fermionic sums
which explicitly involve the $E_8$ root system \cite{WP}. 
This ``fermionic sum = bosonic sum'' expression yields the 
$E_8$ Rogers-Ramanujan identity for the $\chi_{1,1}^{(3,4)}$ Virasoro
character \cite{KKMM}.
The Fourier transform results for the single particle excitations \cite{BNW}
have been inverted to obtain analytic expressions for the excitation
energies of the eight quasiparticles in the Hamiltonian version of the 
model \cite{MO}.

Including the groundstate, the thermodynamic Bethe Ansatz
relied on the input of nine sets of
thermodynamically significant string solutions of the Bethe equations. 
These string types were checked numerically as far as possible at 
criticality \cite{BNW}.
The stability of these strings types away from criticality has been examined
in detail \cite{GNa,GN}. One of the string types, associated with the 
mass $m_4$,
was seen to differ from that proposed in \cite{BNW}. String distributions
were observed up to mass $m_5$, with a number of elementary excitations up
to mass $m_1+m_4$. However, root distributions associated with the
masses $m_6$ to $m_8$ were not observed, presumably because of the inherent
numerical difficulties.  
 
In this paper we explicitly derive the inverse correlation lengths, and 
thus the mass gaps, of the dilute $A_3$ lattice model.\footnote{The result for
the first correlation length has been given in \cite{BS}.} We use the 
string solutions given in \cite{BNW,GN} and apply Baxter's exact perturbative
approach \cite{Baxter}, as used for example in the calculation of 
correlation lengths in the cyclic solid-on-solid (CSOS) model \cite{PB}. 
In particular, we obtain the mass gaps given in (\ref{masses}) 
as criticality is approached.

The outline of the paper is as follows. The dilute $A_L$ lattice model
is defined along with the corresponding Bethe equations in Section 2.
The bulk free energy is derived via the exact perturbation approach
in Section 3. The eigenvalue expression for the leading excitations 
in regime 1 is derived for $L$ odd in Section 4. 
The eigenvalue expressions in regime 2 for $L=3$ associated with the 
eight $E_8$ masses are derived in Section 5. 
These results are collected together in one formula in
Section 6, where the corresponding correlation lengths and masses are 
given.
The paper concludes with a discussion of the results in Section 7.
Some intermediary results are given in the Appendices. 

\section{The dilute A$_L$ lattice model}

\setcounter{equation}{0}

The dilute A$_L$ model
is an exactly solvable, restricted solid-on-solid
model defined on the square lattice. At criticality, the model
can be constructed \cite{WNS,R} from the dilute $O(n)$ 
model \cite{N,WN}. 
Each site of the lattice can take one of $L$ possible
(height) values, subject to the restriction that
neighbouring sites of the lattice either have the
same height, or differ by $\pm 1$.
The off-critical Boltzmann weights of the allowed height 
configurations of an elementary face of the lattice are \cite{WNS}
\begin{eqnarray}
\W{}{a}{a}{a}{a}&=&
\frac{\te(6\la-u)\te(3\la+u)}{\te(6\la)\te(3\la)} 
-\left[\frac{S(a+1)}{S(a)}\frac{\tv(2a\la-5\la)}{\tv(2a\la+\la)}
\right.
\nonumber \\
&\phantom{=}& \left.
+\,\,\frac{S(a-1)}{S(a)}\frac{\tv(2a\la+5\la)}{\tv(2a\la-\la)}\right]
\frac{\te(u)\te(3\la-u)}{\te(6\la)\te(3\la)} \, ,
\nonumber \\
\W{}{a}{a}{a}{a\pm 1}&=&\W{}{a}{a\pm 1}{a}{a}=
\frac{\te(3\la-u)\tv(\pm 2a\la+\la-u)}{\te(3\la)\tv(\pm 2a\la+\la)} \, ,
\nonumber \\ 
\W{}{a\pm 1}{a}{a}{a}&=&\W{}{a}{a}{a\pm 1}{a}
\nonumber \\
&=& \left[\frac{S(a\pm 1)}{S(a)}\right]^{1/2}
\frac{\te(u)\tv(\pm 2a\la-2\la+u)}{\te(3\la)\tv(\pm 2a\la+\la)} \, ,
\nonumber \\ 
\W{}{a}{a\pm 1}{a\pm 1}{a}&=&\W{}{a}{a}{a\pm 1}{a\pm 1}
\nonumber \\ 
&=&\left[\frac{\tv(\pm 2a\la+3\la)\tv(\pm 2a\la-\la)}
           {\tv^2(\pm 2a\la+\la)}\right]^{1/2}
\nonumber \\ 
&\phantom{=}& \times \,\,
\frac{\te(u)\te(3\la-u)}{\te(2\la)\te(3\la)} \, , 
\nonumber \\
\W{}{a}{a\mp 1}{a}{a\pm 1}&=&
\frac{\te(2\la-u)\te(3\la-u)}{\te(2\la)\te(3\la)} \, ,
\nonumber \\ 
\W{}{a\pm 1}{a}{a\mp 1}{a}&=&
-\left[\frac{S(a-1)S(a+1)}{S^2(a)}\right]^{1/2}
\frac{\te(u)\te(\la-u)}{\te(2\la)\te(3\la)} \, ,
\nonumber \\ 
\W{}{a\pm 1}{a}{a\pm 1}{a}&=&
\frac{\te(3\la-u)\te(\pm 4a\la+2\la+u)}{\te(3\la)\te(\pm 4a\la+2\la)}
\nonumber \\
&\phantom{=}& + \,\, \frac{S(a\pm 1)}{S(a)}
\frac{\te(u)\te(\pm 4a\la-\la+u)}{\te(3\la) \te(\pm 4a\la+2\la)}
\nonumber \\ 
&=& \frac{\te(3\la+u)\te(\pm 4a\la-4\la+u)}
{\te(3\la)\te(\pm 4a\la-4\la)}
\nonumber \\ 
&\phantom{=}& + \,\,
\left[\frac{S(a\mp 1)}{S(a)}\frac{\te(4\la)}{\te(2\la)}
-\frac{\tv(\pm 2a\la-5\la)}{\tv(\pm 2a\la+\la)} \right]
\nonumber \\ 
&\phantom{=}& \times \,\,
\frac{\te(u)\te(\pm 4a\la-\la+u)}{\te(3\la) \te(\pm 4a\la-4\la)}. 
\label{Bweights}
\end{eqnarray}

The crossing factors $S(a)$ are defined by
\begin{equation}
S(a)  =  (-1)^{\displaystyle a} \;\frac{\vartheta_1({4a\lambda})}{
           \vartheta_4({2a\lambda})}
\end{equation}
and $\vartheta_1({u})$, $\vartheta_4({u})$ are standard elliptic
    theta functions of nome $p$,
\begin{eqnarray}
\vartheta_1(u)&=&\vartheta_1(u,p)=2p^{1/4}\sin u\:
  \prod_{n=1}^{\infty} \left(1-2p^{2n}\cos
   2u+p^{4n}\right)\left(1-p^{2n}\right) , \label{theta1}  \\
\vartheta_4(u)&=&\vartheta_4(u,p)=\prod_{n=1}^{\infty}\left(
 1-2p^{2n-1}\cos2u+p^{4n-2}\right)\left(1-p^{2n}\right). 
  \label{theta4}
\end{eqnarray}

In the above weights the range of the spectral parameter $u$
and the variable $\lambda$ are given by $0<u< 3\lambda$ with
\begin{equation}
\lambda =   \frac{s}{r} \; \pi ,
\end{equation}
where $r=4(L+1)$, with $s=L$ in regime 1 and $s=L+2$ 
in regime 2. We do not consider the other regimes here.
The thermal Ising point occurs in regime 1 with $L=2$
and the magnetic Ising point occurs in regime 2 with $L=3$.

The row transfer matrix of the dilute A models is
defined on a periodic strip of width $N$ as 
\begin{equation}
T_{\{a\}}^{\{b\}} = \prod_{j=1}^{N}
\W{}{a_j}{a_{j+1}}{b_{j+1}}{b_j} , 
\end{equation}
where $\{a\}$ is an admissible path of heights and
$a_{N+1} =a_1$, $b_{N+1} = b_1$. For convenience we
take $N$ even.

The eigenvalues of the transfer matrix are \cite{BNW,ZPG,Zh}
\begin{eqnarray}
\Lambda(u) &=& \omega \left[
\frac{\te(2\lambda-u)\;\te(3\lambda-u)}{\te(2\lambda)\;\te(3\lambda)}
\right]^N
\prod_{j=1}^N
\frac{\te(u-u_j+\lambda)}{\te(u-u_j-\lambda)}
\nonumber \\
&&+ \left[
\frac{\te(u)\;\te(3\lambda-u)}{\te(2\lambda)\;\te(3\lambda)}
\right]^N
\prod_{j=1}^N
\frac{\te(u-u_j) \; \te(u-u_j-3\lambda)}
     {\te(u-u_j-\lambda) \; \te(u-u_j-2 \lambda)} 
\nonumber\\
&& + \, \omega^{-1}
\left[
\frac{\te(u)\;\te(\lambda-u)}{\te(2\lambda)\;\te(3\lambda)}
\right]^N
\prod_{j=1}^N
\frac{\te(u-u_j-4\lambda)}{\te(u-u_j-2\lambda)} , 
\label{eigs}
\end{eqnarray}
where the $N$ roots $u_j$ are given by the Bethe equations 
\begin{equation}
\omega \left[
\frac{\te(\lambda-u_j)}{\te(\lambda+u_j)}\right]^{N} =
-\prod_{k=1}^{N}
\frac{\te(u_j - u_k - 2\lambda) \; \te(u_j - u_k + \lambda) }
     {\te(u_j - u_k + 2\lambda) \; \te(u_j - u_k - \lambda) }
\label{BAE}
\end{equation}
and $\omega=\exp(\ib \pi \ell/(L+1))$ for $\ell=1,\ldots,L$.

There are several methods at hand to calculate the correlation length.
Here we apply the perturbative approach initiated by 
Baxter \cite{Baxter,PB}.
For $L$ odd this involves perturbing away from the high magnetic 
field limit at $p=1$. We thus introduce the conjugate variables
\begin{equation}
w = \e^{-2\pi u/\epsilon} \quad \mbox{and} \quad 
x = \e^{- \pi^2/r \epsilon},
\end{equation}  
where nome $p=\e^{-\epsilon}$.
The relevant conjugate modulus transformations are
\begin{eqnarray}
\te(u,p) &=& \left( {\pi \over \epsilon} \right)^{1/2} 
             \e^{-(u-\pi/2)^2/\epsilon} \,
              E(w,q^2) ,  \\
\tv(u,p) &=& \left( {\pi \over \epsilon} \right)^{1/2} 
             \e^{-(u-\pi/2)^2/\epsilon} \,
              E(-w,q^2) , 
\end{eqnarray}
where $q=\e^{-\pi^2/\epsilon}$ and 
\begin{equation}
E(z,p) = \prod_{n=1}^{\infty} (1-p^{n-1} z)(1-p^n z^{-1})(1-p^n).
\end{equation}

In the ordered limit ($p\to 1$ with $u/\epsilon$ fixed) the 
Boltzmann weights for $L$ odd reduce to
\begin{equation}
\W{}{a}{b}{c}{d} \sim w^{H(d,a,b)} \, \delta_{a,c} \, .
\end{equation}
The function $H(d,a,b)$ is given explicity in \cite{WPSN},
being required for the calculation of the local height
probabilities. In this limit the row transfer matrix 
eigenspectra breaks up into a number of distinct bands labelled by
integer powers of $w$. In regime 1 there are $\case{1}{2}(L+1)$ 
ground states and in regime 2 there are
$\case{1}{2}(L-1)$ ground states, each with eigenvalue $\Lambda_0=1$. 
The bands of excitations are relevant to the calculation of the
correlation lengths. 

The number of states in the $w$ band is $\case{1}{2}(L-1) N$ in
regime 1 and $\case{1}{2}(L-3) N$ in regime 2. These correspond to
introducing in all but one of the ground state paths $\{a\}$
a single non-ground state height, in any position. 
In particular, note that there are {\em no} excitations in
the $w$ band for $L=3$ in regime 2. Thus for the magnetic Ising model
the leading excitations are in the $w^2$ band. These are harder to
count, arising from a variety of both single and multiple 
deviations from ground state paths.
However, we observe numerically that (apart from when $N=2$) 
there are $4N$ states in this $w^2$ band.     

In our numerical investigation of the transfer matrix eigenspectrum
we associate a given value of $\ell$ with each eigenvalue
by comparing the eigenspectrum at criticality ($p=0$) 
with the eigenspectrum of the corresponding O($n$) loop model 
for finite $N$.\footnote{Strictly speaking we compare with the
eigenspectrum of the corresponding vertex model with seam $\omega$.} 
Each eigenvalue can then be tracked to the ordered limit. In this way
the band of largest eigenvalues is seen to have the values
$\ell = 1,\ldots,\case{1}{2}(L+1)$ in regime 1 and
$\ell = 1,\ldots,\case{1}{2}(L-1)$ in regime 2, i.e. one
value of $\ell$ for each ground state.

Setting $w_j = \e^{-2\pi u_j/\epsilon}$, the eigenvalues (\ref{eigs}) 
can be written
\begin{eqnarray}
\Lambda(w) &=& \omega \left[
\frac{E(x^{4s}/w,x^{2r})\;E(x^{6s}/w,x^{2r})}
     {E(x^{4s},x^{2r})\;E(x^{6s},x^{2r})}
\right]^N
\prod_{j=1}^N w_j^{1 - 2s/r} \,
\frac{E(x^{2s} w /w_j,x^{2r})}{E(x^{2s} w_j/w,x^{2r})}
\nonumber \\
&\phantom{=}& +\, \left[ 
\frac{x^{2s}}{w}
\frac{E(w,x^{2r})\;E(x^{6s}/w,x^{2r})}
     {E(x^{4s},x^{2r})\;E(x^{6s},x^{2r})}
\right]^N
\nonumber \\
&\phantom{=}& \qquad \times 
\prod_{j=1}^N w_j \,
\frac{E(w/w_j,x^{2r})\;E(x^{6s} w_j /w,x^{2r})}
     {E(x^{2s} w_j/w,x^{2r})\;E(x^{4s} w_j /w,x^{2r})}
\nonumber\\
&\phantom{=}& +\, \omega^{-1}
\left[ x^{2s} \,
\frac{E(w,x^{2r})\;E(x^{2s}/w,x^{2r})}
     {E(x^{4s},x^{2r})\;E(x^{6s},x^{2r})}
\right]^N \prod_{j=1}^N w_j^{2s/r}
\frac{E(x^{8s} w_j /w,x^{2r})}{E(x^{4s} w_j /w,x^{2r})} .
\nonumber \\
\label{eigsc}
\end{eqnarray}
The Bethe equations (\ref{BAE}) are now 
\begin{eqnarray}
\omega \left[ w_j \, 
\frac{E(x^{2s}/w_j,x^{2r})}{E(x^{2s} w_j,x^{2r})}
\right]^N &&\nonumber \\
= - \prod_{k=1}^N w_k^{2s/r} &&
\frac{E(x^{2s} w_j/w_k,x^{2r}) \, E(x^{4s} w_k/w_j,x^{2r})}
     {E(x^{2s} w_k/w_j,x^{2r}) \, E(x^{4s} w_j/w_k,x^{2r})} .
\label{BAEc}
\end{eqnarray}
We are now ready to investigate the ordered limit.

\section{Free energy}
 
\setcounter{equation}{0}
 
The calculation of the largest eigenvalue in the thermodynamic limit
proceeds from the $x\to0$ limit with $w$ fixed in a similar manner
to that for the eight-vertex \cite{Baxter} and CSOS \cite{PB} models.
We make repeated use of the properties 
\begin{equation} 
E(z,p)=E(p/z,p)=-zE(z^{-1},p) .
\end{equation}
Assuming that the roots $w_j$ are on the unit circle, as observed
in our numerical calculations, only the first term in the eigenvalue 
expression (2.14) survives in this limit, with
\begin{equation}
\Lambda_0\sim \omega (w_1 \ldots w_N)^{(r-2s)/r} .
\end{equation}
The Bethe equation (\ref{BAEc}) gives
\begin{equation}
w^N +\omega^{-1} (w_1 \ldots w_N)^{2s/r}=0 \,,
\end{equation}
where we write $w=w_j$. We consider this equation for all complex
$w$ and equate the left hand side to
$
\prod_{j=1}^N(w-w_j)
$
so that 
\begin{equation}
(w_1 \ldots w_N)=\omega^{-1} (w_1 \ldots w_N)^{2s/r}\Rightarrow
(w_1 \ldots w_N)^{(r-2s)/r} = \omega^{-1} \label{roots} .
\end{equation}
Recall that we have taken $N$ even. Thus from (3.2) the
largest eigenvalue is $\Lambda_0=1$ in the limit $x \to 0$.
Further, we expect the product of the roots to obey (\ref{roots}) 
away from $x=0$.
Each of the degenerate ground states has a different root
distribution $\{w_j\}$ on the unit circle, depending on $\ell$.

To perturb about $x=0$, we define the auxiliary functions
\begin{eqnarray}
A(z)&=&\prod_{k=0}^{\infty}(1-x^{2rk}z)^N ,\label{defA}\\
F_0(w)&=&\prod_{j=1}^N  \prod_{k=0}^{\infty} (1-x^{2rk}w/w_j) ,\\
G_0(1/w)&=&\prod_{j=1}^N  \prod_{k=1}^{\infty} (1-x^{2rk}w_j/w), 
\end{eqnarray}
where $A(z)$ is known, and $F_0(w)$ and $G_0(1/w)$, which depend upon
the $w_j$, are the unknowns to be found.
Then the Bethe equation (\ref{BAEc}) may be written as
\begin{eqnarray}
w^N && \frac{A(x^{2s}/w)G_0(1/x^{4s}w)G_0(1/x^{2r-2s}w)
}
{A(x^{2r-2s}/w)
G_0(1/x^{2s}w)G_0(1/x^{2r-4s}w)}+ 
\nonumber \\
&&(w_1 \ldots w_N)
\frac{A(x^{2s}w)F_0(x^{2s}w)F_0(x^{2r-4s}w)}
{A(x^{2r-2s}w)F_0(x^{4s}w)F_0(x^{2r-2s}w)}=0 ,
\end{eqnarray}
which is again an $N\th$ order equation with roots $w_1, \ldots w_N$,
so that the left hand side may be equated to 
\begin{equation}
\prod_{j=1}^N(w-w_j)=\left\{\begin{array}{ll}
w^N\prod_{j=1}^N(1-w_j/w),& \quad \mbox{$w$ large}, \\
(w_1 \ldots w_N)\prod_{j=1}^N(1-w/w_j), &\quad \mbox{$w$ small} .
\end{array} \right. 
\end{equation}
These products can be expressed in terms of the auxiliary functions,
\begin{equation}
\prod_{j=1}^N(1-w/w_j)=\frac{F_0(w)}{F_0(x^{2r}w)}, \qquad
\prod_{j=1}^N(1-w_j/w)=\frac{G_0(1/x^{2r}w)}{G_0(1/w)}.
\end{equation}
If we then further define 
\begin{equation}
\F_0(w) =\frac{F_0(w)}{F_0(x^{2r-4s}w)},
\qquad \G_0(1/w)=\frac{G_0(1/w)}{G_0(1/x^{2r-4s}w)},
\end{equation}
we obtain two equations,
\begin{eqnarray}
\F_0 (w)&=&\frac{A(x^{2s}w)}{A(x^{2r-2s}w)} \frac{\F_0 (x^{2s}w)}
{\F_0 (x^{4s}w)},\nonumber\\
\G_0(1/w)&=&\frac{A(x^{2r+2s}/w)}{A(x^{6s}/w)} \frac{\G_0 (x^{2s}/w)}
{\G_0(x^{4s}/w)},
\end{eqnarray}
by equating the dominant terms for $|w|>1$ and for $|w|<1$.
These equations can be solved by iteration to give
\begin{eqnarray}
\F_0(w)&=&\prod_{m=0}^{\infty}
\frac{A(x^{(12m+2)s}w)A(x^{(12m+4)s}w)}
{A(x^{(12m+8)s}w)A(x^{(12m+10)s}w)} \nonumber\\
&\phantom{=}& \quad \times \frac{A(x^{(12m+4)s+2r}w)A(x^{(12m+6)s+2r}w)}
{A(x^{(12m-2)s+2r}w)A(x^{12sm+2r}w)}, \\ 
\G_0(1/w)&=&\prod_{m=0}^{\infty}
\frac{A(x^{(12m+12)s}/w)A(x^{(12m+14)s}/w)}
{A(x^{(12m+6)s}/w)A(x^{(12m+8)s}/w)}  \nonumber\\
&\phantom{=}& \quad \times
\frac{A(x^{(12m+2)s+2r}/w)A(x^{(12m+4)s+2r}/w)}
{A(x^{(12m+8)s+2r}/w)A(x^{(12m+10)s+2r}/w)}. 
\end{eqnarray}

The eigenvalue (\ref{eigsc}) can be expressed in terms
of the auxiliary functions as
\begin{eqnarray}
\Lambda_0&=&
\frac{A(x^{4s}/w)A(x^{6s}/w)A(x^{2r-4s}w)A(x^{2r-6s}w)}
{A(x^{4s})A(x^{6s})A(x^{2r-4s})A(x^{2r-6s})}
\F_0(x^{2s}w)\G_0(1/x^{2s}w) \nonumber \\
&&+\frac{A(w)A(x^{6s}/w)A(x^{2r}/w)A(x^{2r-6s}w)}
{A(x^{4s})A(x^{6s})A(x^{2r-4s})A(x^{2r-6s})} 
\frac{\F_0(w)\G_0(1/w)}{\F_0(w/x^{2s})\G_0(x^{2s}/w)} \nonumber\\
&&+\frac{A(w)A(w/x^{2s})A(x^{2r}/w)A(x^{2r+2s}/w)}
{A(x^{4s})A(x^{6s})A(x^{2r-4s})A(x^{2r-6s})} 
\frac{1}{\F_0(w/x^{4s})\G_0(x^{4s}/w)}, 
\end{eqnarray}
where we have made use of (\ref{roots}).
When the above solutions for $\F_0(w)$ and $\G_0(1/w)$ are used, 
we can show that all three terms
are identical, so that we finally obtain
\begin{eqnarray}
\Lambda_0/3&=&
\frac{A(x^{4s}/w)A(x^{6s}/w)A(x^{2r-4s}w)A(x^{2r-6s}w)}
{A(x^{4s})A(x^{6s})A(x^{2r-4s})A(x^{2r-6s})} \nonumber\\
&& \prod_{m=0}^{\infty}
\frac{A(x^{(12m+4)s}w)A(x^{(12m+6)s}w)}{A(x^{(12m+10)s}w)A(x^{(12m+12)s}w)}
\nonumber\\ && \quad \times
\frac{A(x^{(12m+6)s+2r}w)A(x^{(12m+8)s+2r}w)}
{A(x^{12ms+2r}w)A(x^{(12m+2)s+2r}w)} \nonumber\\
&& \quad \times
\frac{A(x^{12ms+2r}/w)A(x^{(12m+2)s+2r}/w)}{A(x^{(12m+6)s+2r}/w)
A(x^{(12m+8)s+2r}/w)}
\nonumber\\ && \quad \times
\frac{A(x^{(12m+10)s}/w)A(x^{(12m+12)s}/w)}
{A(x^{(12m+4)s}/w)A(x^{(12m+6)s}/w)} . 
\end{eqnarray}
The factor of 3 will not survive the thermodynamic limit. 
Using the identity
\begin{equation}
\frac{1}{N} \log A(w)=- \sum_{k=1}^{\infty} \frac{w^k}{k(1-x^{2rk})}
\end{equation}
and defining the free energy per site as $f = -N^{-1} \log \Lambda_0$
we are able to write 
\begin{equation}
f = -
\sum_{k=1}^{\infty}
\frac{(1-w^k)(1-x^{6sk}w^{-k})(x^{4sk}+x^{(2r-6s)k})(1+x^{2sk})}
{k(1-x^{2rk})(1+x^{6sk})}.
\end{equation}
Our final result is thus
\begin{equation}
f =  4 \sum_{k=1}^{\infty}
\frac{ \c [(5 \lambda-\pi) \pi k/\epsilon] \,
\c (\pi \lambda k/\epsilon) \,
\s (\pi u k/\epsilon) \,
\s [(3\lambda-u)\pi k/\epsilon]}
{k\, \s (\pi^2 k/\epsilon) \,
\c (3 \pi \lambda k/\epsilon)}. \label{fen}
\end{equation}
This is in agreement with the previous calculations via the inversion
relation method \cite{WNS,WPSN}. The singular behaviour follows as 
\begin{equation}
f \sim \left\{ \begin{array}{ll}
                 p^{1 + 1/\delta} & \quad \mbox{$L$ odd}, \\
                 p^{2 - \alpha}   & \quad \mbox{$L$ even},
             \end{array} \right. \quad \mbox{as} \quad p \to 0,
\end{equation}
where the exponents are given by \cite{WNS,WPSN}
\begin{equation}
\begin{array}{lll}
\delta = \frac{3L}{L+4}, & \,\, \alpha = \frac{2(L-2)}{3 L} & 
                                \quad \mbox{regime 1}, \\
\delta = \frac{3(L+2)}{L-2}, & \,\, \alpha = \frac{2(L+4)}{3(L+2)} & 
                                \quad \mbox{regime 2}.
\end{array}
\end{equation}
These include the Ising values $\alpha=0$ for $L=2$ in regime 1 and
$\delta=15$ for $L=3$ in regime 2.

\section{Excitations in regime 1}
 
\setcounter{equation}{0}

In regime 1, we observe numerically that the leading eigenvalue in the $w$ band has 
$\ell=\case{1}{2}(L+1)+1$. The corresponding root distribution has
$N-1$ roots on the unit circle and a 1-string excitation
located exactly at $w_N = - x^r$. In general we 
assume that the 1-string excitations are located at $w_N=bx^r$
with $|b| \sim 1$ and the remainder on the unit circle, i.e.   
$w_j=a_j$ for $j=1, \ldots N-1$.
The Bethe equations split into two sets, the first of which is 
\begin{eqnarray}
\omega \left[a_k \frac{E(x^{2s}/a_k)}{E(x^{2s}a_k)}\right]^N
&=&-(A_{N-1}b)^{2s/r}
\frac{E(x^{r-2s}b/a_k)E(x^{r-4s}a_k/b)}{E(x^{r-2s}a_k/b)E(x^{r-4s}b/a_k)}
\nonumber\\
&&\quad \times \prod_{j=1}^{N-1}
\frac{E(x^{2s}a_k/a_j)E(x^{4s}a_j/a_k)}{E(x^{2s}a_j/a_k)E(x^{4s}a_k/a_j)},
\end{eqnarray} 
for $k=1, \ldots, N-1$, where we have defined $\prod_{j=1}^{m}a_j=A_m$.
For $k=N$ the other equation is
\begin{equation}
\omega \left[\frac{E(x^{r-2s}b)}{E(x^{r-2s}/b)}\right]^N
=-(A_{N-1}b)^{2s/r}\prod_{j=1}^{N-1}
\frac{E(x^{r-2s}a_j/b)E(x^{r-4s}b/a_j)}{E(x^{r-2s}b/a_j)E(x^{r-4s}a_j/b)}.
\label{2nd}
\end{equation} 
Taking the limit $x \to 0$, we obtain 
\begin{eqnarray}
a^N + \omega^{-1} (A_{N-1}b)^{2s/r}&=&0, \label{e1}\\
- \omega^{-1} (A_{N-1}b)^{2s/r}&=&1, \label{e2}
\end{eqnarray}
which taken together give $a^N=1$, where $a=a_k$. This equation is of order
$N$, but recall there are only $N-1$ of the $a_k$'s. The extra degree of this
equation defines a hole at $a_N$.
Further, from (\ref{e1}) and (\ref{e2}) we also have
\begin{equation}
 \omega^{-1} (A_{N-1}b)^{2s/r}=A_N=A_{N-1}a_N=-1.\label{hole}
\end{equation}
To perform the perturbation, we define in addition to $A(w)$, 
\begin{eqnarray}
X(w)&=&\prod_{k=0}^{\infty}(1-x^{2rk}w/b), \label{defX}\\
Y(1/w)&=&\prod_{k=1}^{\infty}(1-x^{2rk}b/w), \label{defY}\\
R(w)&=&\prod_{k=0}^{\infty}(1-x^{2rk}w/a_N), \\
S(1/w)&=&\prod_{k=1}^{\infty}(1-x^{2rk}a_N/w), 
\end{eqnarray}
and the functions we will need to find,
\begin{eqnarray}
F(w)&=&\prod_{j=1}^N \prod_{k=0}^{\infty} (1-x^{2rk}w/a_j), \nonumber\\
G(1/w)&=&\prod_{j=1}^N \prod_{k=1}^{\infty} (1-x^{2rk}a_j/w).
\end{eqnarray}
The Bethe equation for $k=1, \ldots, N-1$ can now be written
\begin{eqnarray}
&& a^N \, \frac{A(x^{2s}/a)}{A(x^{2r-2s}/a)} 
\frac{Y(1/x^{r+4s}a)Y(1/x^{r-2s}a)}{Y(1/x^{r+2s}a)Y(1/x^{r-4s}a)}
\nonumber \\ && \quad \times
\frac{S(1/x^{2s}a)S(1/x^{2r-4s}a)}{S(1/x^{4s}a)S(1/x^{2r-2s}a)} 
\frac{G(1/x^{4s}a)G(1/x^{2r-2s}a)}{G(1/x^{2s}a)G(1/x^{2r-4s}a)}
\nonumber \\
&& + \, \, \omega^{-1}(A_{N-1}b)^{2s/r}
\frac{A(x^{2s}a)}{A(x^{2r-2s}a)} 
\frac{X(x^{r-4s}a)X(x^{r+2s}a)}{X(x^{r-2s}a)X(x^{r+4s}a)} \nonumber \\
&& \quad \times 
\frac{R(x^{4s}a)R(x^{2r-2s}a)}{R(x^{2s}a)R(x^{2r-4s}a)} 
\frac{F(x^{2s}a)F(x^{2r-4s}a)}{F(x^{4s}a)F(x^{2r-2s}a)}=0.
\end{eqnarray}
As before, we equate the lhs of this to
\begin{equation}
\prod_{j=1}^N(a-a_j)= \left\{ \begin{array}{ll}
a^N G(1/x^{2r}a)/G(1/a),& \quad \mbox{$a$ large}, \\
A_N F(a)/F(x^{2r}a),& \quad \mbox{$a$ small}.
\end{array} \right.
\end{equation}
We are thus led to define
\begin{equation}
\F(a)=\frac{F(a)}{F(x^{2r-4s}a)}, 
\quad \G(1/a)=\frac{G(1/a)}{G(1/x^{2r-4s}a)},
\label{fgdef}
\end{equation}
which satisfy the recurrences
\begin{eqnarray}
\F(a)&=&\frac{A(x^{2s}a)}{A(x^{2r-2s}a)}
\frac{X(x^{r-4s}a)X(x^{r+2s}a)}{X(x^{r-2s}a)X(x^{r+4s}a)} \nonumber \\
&\phantom{=}& \quad \times 
\frac{R(x^{4s}a)R(x^{2r-2s}a)}{R(x^{2s}a)R(x^{2r-4s}a)}
\frac{\F(x^{2s}a)}{\F(x^{4s}a)}, \nonumber\\
\G(1/a)&=&\frac{A(x^{2r+2s}/a)}{A(x^{6s}/a)}
\frac{Y(1/x^{r-2s}a)Y(1/x^{r-8s}a)}{Y(1/x^{r}a)Y(1/x^{r-6s}a)}
\nonumber \\ &\phantom{=}& \quad \times
\frac{S(1/a)S(1/x^{2r-6s}a)}{S(x^{2s}/a)S(1/x^{2r-8s}a)}
\frac{\G(x^{2s}/a)}{\G(x^{4s}/a)}.
\end{eqnarray}

Their solutions are
\begin{eqnarray}
\F(a)&=&\F_0(a)\frac{X(x^{r-4s}a)R(x^{2r}a)}{X(x^r a) R(x^{2r-4s}a)}
\nonumber\\ &\phantom{=}& \quad \times
\prod_{m=0}^{\infty} \frac{(1-x^{(12m+6)s}a/a_N)(1-x^{(12m+8)s}a/a_N)}
{(1-x^{(12m+2)s}a/a_N)(1-x^{(12m+12)s}a/a_N)}, \label{solF}\\
\G(1/a)&=&\G_0(1/a)\frac{Y(1/x^{r-4s}a)S(1/a)}{Y(1/x^r a) S(x^{4s}/a)}
\nonumber\\ &\phantom{=}& \quad \times
\prod_{m=0}^{\infty} \frac{(1-x^{(12m+6)s}a_N/a)(1-x^{(12m+16)s}a_N/a)}
{(1-x^{(12m+10)s}a_N/a)(1-x^{(12m+12)s}a_N/a)}, \label{solG}
\end{eqnarray}
where $\F_0$ and $\G_0$ are the functions used in the previous Section.

We can now construct the eigenvalue, which in terms of the auxiliary
functions reads
\begin{eqnarray}
\Lambda &=& A_{N-1}w 
\frac{A(x^{2r-4s}w)A(x^{2r-6s}w)A(x^{4s}/w)A(x^{6s}/w)}
{A(x^{4s})A(x^{6s})A(x^{2r-4s})A(x^{2r-6s})} \nonumber\\
&\phantom{=}& \times 
\frac{R(x^{2r-2s}w)S(1/x^{2r-2s}w)}{R(x^{2s}w)S(1/x^{2s}w)} \nonumber\\
&\phantom{=}& \times
\frac{X(x^{r+2s}w)Y(1/x^{r+2s}w)}{X(x^{r-2s}w)Y(1/x^{r-2s}w)}
\F(x^{2s}w)\G(1/x^{2s}w) \nonumber \\
&+&
\frac{A(w)A(x^{2r-6s}w)A(x^{2r}/w)A(x^{6s}/w)}
{A(x^{4s})A(x^{6s})A(x^{2r-4s})A(x^{2r-6s})} \nonumber\\
&\phantom{=}& \times
\frac{R(x^{2r-4s}w)R(w/x^{2s})}{R(w)R(x^{2r-6s}w)}
\frac{S(1/x^{2r-4s}w)S(x^{2s}/w)}{S(1/w)S(1/x^{2r-6s}w)} \nonumber\\
&\phantom{=}& \times
\frac{X(x^{r-6s}w)X(w/x^{r})}{X(x^{r-4s}w)X(w/x^{r+2s})}
\frac{Y(1/x^{r-6s}w)Y(x^{r}/w)}{Y(1/x^{r-4s}w)Y(x^{r+2s}/w)} \nonumber\\
&\phantom{=}& \times
\frac{\F(w)\G(1/w)}{\F(w/x^{2s})\G(x^{2s}/w)} \nonumber \\
&+&
\frac{x^{2s-r}}{A_{N-1}b}
\frac{A(w)A(x^{2r}/w)A(w/x^{2s})A(x^{2r+2s}/w)}
{A(x^{4s})A(x^{6s})A(x^{2r-4s})A(x^{2r-6s})} \nonumber \\
&\phantom{=}& \times
\frac{R(w/x^{4s})S(x^{4s}/w)}{R(x^{2r-8s}w)S(1/x^{2r-8s}w)} \nonumber \\
&\phantom{=}& \times 
\frac{X(x^{r-8s}w)Y(1/x^{r-8s}w)}{X(w/x^{r+4s})Y(x^{r+4s}/w)}
\frac{1}{\F(w/x^{4s})\G(x^{4s}/w)} \,. 
\end{eqnarray}
Making use of the solutions (\ref{solF}) and (\ref{solG}), the identities 
in Appendix A, and (\ref{hole}) on the prefactors, we can show that this 
simplifies to
\begin{equation}
\Lambda/\Lambda_0=wA_{N-1}\frac{E(x^{8s}w/a_N, x^{12s})E(x^{10s}w/a_N,x^{12s})}
{E(x^{4s}w/a_N, x^{12s})E(x^{2s}w/a_N,x^{12s})}.
\label{genreg1}
\end{equation}
If both the hole $a_N$, and $b$ are equal to $-1$, then we have
\begin{equation}
\frac{\Lambda}{\Lambda_0} = w\, 
\frac{E(-x^{2s}/w,x^{12s})\,E(-x^{4s}/w,x^{12s})}
     {E(-x^{2s}\,w,x^{12s})\,E(-x^{4s}\,w,x^{12s})} \,.
\label{reg1}
\end{equation}
At the isotropic point $w = x^{3s}$ ($u=3\lambda/2$) this reduces to
\begin{equation}
\frac{\Lambda}{\Lambda_0} = x^{s}\,
\frac{E^2(-x^s,x^{12s})}
     {E^2(-x^{5s},x^{12s})} =
\left[
\frac{\tv(\frac{\pi}{12},p^{\pi/6\lambda})}
     {\tv(\frac{5\pi}{12},p^{\pi/6\lambda})}
\right]^2 .
\label{gapreg1}
\end{equation}
We can also express the second Bethe equation (\ref{2nd}) 
in terms of the auxiliary functions. This yields an equation for the
parameters $b$ and $a_N$ of the form
\begin{eqnarray}
&&\frac{A(x^{r-2s}b)A(x^{r+2s}/b)}{A(x^{r-2s}/b)A(x^{r+2s}b)}=\nonumber \\
&& \qquad \phantom{\times}
\frac{R(x^{r-2s}b)R(x^{r+4s}b)}{R(x^{r-4s}b)R(x^{r+2s}b)}
\frac{S(1/x^{r-2s}b)S(1/x^{r+4s}b)}{S(1/x^{r-4s}b)S(1/x^{r+2s}b)}\nonumber \\
&& \qquad \times 
\frac{F(x^{r-4s}b)F(x^{r+2s}b)}{F(x^{r-2s}b)F(x^{r+4s}b)}
\frac{G(1/x^{r-4s}b)G(1/x^{r+2s}b)}{G(1/x^{r-2s}b)G(1/x^{r+4s}b)}.
\end{eqnarray}
Since this cannot be expressed in terms of $\F$ and $\G$, we have
to re-construct
\begin{equation}
F(a)=\prod_{\ell=0}^{\infty}\F(x^{(2r-4s)\ell}a),
\qquad G(1/a)=\prod_{\ell=0}^{\infty}\G(1/x^{(2r-4s)\ell}a),
\end{equation}
from the definitions (\ref{fgdef}).
Making use of identities listed in Appendix A
we finally obtain a result,
\begin{equation}
\frac{E(x^r a_N/b, x^{2r-4s})}{E(x^r b/a_N, x^{2r-4s})} = 1,
\end{equation}
in which neither the original nome $x^{2r}$
nor $x^{12s}$ appears. This is obviously satisfied by $a_N=b$.

\section{Excitations in regime 2 for $L=3$}
 
\setcounter{equation}{0}
 
For $L=3$ in regime 2 extensive numerical investigations of the 
Bethe equations (\ref{BAE}) have led to a conjecture 
for the thermodynamically significant strings \cite{BNW,GNa,GN}.
For example, the leading excitation $\Lambda_1$, 
corresponding to mass $m_1$, is a 2-string with $\ell=2$. 
However, this state is originally a 1-string for small $p$ and small $N$. 
Such behaviour has been discussed in \cite{GN}. By tracking this state 
numerically with increasing $p$ we see that
this 2-string is exactly located at $-x^{\pm 11}$ in the limit $p=1$.
Moreover, it lies in the $w^2$ band of excitations.
The string structure for the other eigenvalues ($\Lambda_j$) or
masses ($m_j$) used in the thermodynamic Bethe Ansatz approach were  
checked numerically as far as possible at criticality \cite{BNW}.
The stability of these strings types away from criticality has been 
examined in detail \cite{GNa,GN}.
The conjectured string positions are listed in Table 1.
With the exception of mass $m_4$, these are the string positions
given in \cite{BNW}. Rather than the 5-string, $\pm 8, \pm 12, 16$, 
proposed in \cite{BNW} we consider the 7-string found in \cite{GN}. 
In the latter study, string distributions for elementary excitations 
up to and including mass $m_5$ were located, 
including a number of composite excitations up to mass $m_1+m_4$. 
However, the root distributions associated with the
masses $m_6$ to $m_8$ were not observed, presumably because of
the inherent numerical difficulties. 

Here we show that each of the tabulated sets of strings provides
a consistent solution in the ordered and large $N$ limits. 
However, this is also true for the proposed 5-string for mass $m_4$. 
According to our calculations, both the 5-string and the 7-string 
distributions yield the same mass $m_4$. We shall return to this point 
in the discussion in Section 7. 

\begin{table}[ht]
\caption{String positions and corresponding eigenvalue bands 
for the eight elementary masses $m_j$ of the dilute $A_3$ model
in regime 2. The strings are in units of $\pi \ib /32$.
\vskip 5mm
}
\centerline{
\begin{tabular}{||c|c|c||}
\hline
$j$ &  string positions & band \\
\hline
1   &       $\pm11$  & 2 \\
2   &       $\pm 5, \pm 15$ & 2 \\
3   &       $\pm 10, \pm 12$ & 3 \\
4   &       $\pm 1, \pm 7, \pm 13, 16$ & 3 \\
5   &       $\pm 9, \pm 11, \pm 13$ & 4 \\
6   &       $\pm 6, \pm 10, \pm 14, 16$ & 4 \\
7   &       $\pm 8, \pm 10, \pm 12, \pm 14$ & 5 \\
8   &       $\pm 7, \pm 9, \pm 11, \pm 13, \pm 15$ & 6 \\
\hline
\end{tabular}}
\vskip 1cm
\end{table}

The calculations in this Section are necessarily complicated. The
reader not interested in the specific details may wish to skip to
Sections 6 and 7, where the results obtained are summarised and
discussed.

\subsection{Mass $m_1$}
We begin the perturbation argument with the structure
$w_j=a_j$ for $j=1, \ldots, N-2$ with $w_{N-1}=b_1x^{-11}$ and
$w_N=b_2x^{11}$. However, from the Bethe equations for $k=N-1$ and $k=N$
(see Appendix B) we can show that $b_1=b_2=b$. The Bethe equation for the 
other roots $a_k=a$ is then
\begin{eqnarray}
\lefteqn{-\omega \left[a\frac{E(x^{10}/a)}{E(x^{10}a)}\right]^N=} \nonumber \\
&&(A_{N-2}b^2)^{5/8}\frac{a^2}{b^2}
\frac{E(x^{9}b/a)E(x^{11} b/a)}{E(x^{9}a/b)E(x^{11} a/b)} 
\prod_{j=1}^{N-2} \frac{E(x^{10}a/a_j)E(x^{20}a_j/a)} 
{E(x^{10}a_j/a)E(x^{20}a/a_j)}, \label{bethe1.k}
\end{eqnarray}
where as before, $\prod_{j=1}^ma_j=A_m$.
In the $x \to 0$ limit this gives the equation
\begin{equation}
a^{N-2}+\frac{1}{\omega} (A_{N-2}b^2)^{5/8}/b^2=0,
\end{equation}
which is an $(N-2)\th$ order equation for $N-2$ zeros, so that,
unlike in regime 1, there are no holes in this case. Equating this 
as usual with $\prod_{j=1}^{N-2}(a-a_j)$, we obtain
\begin{equation}
 \frac{1}{\omega} (A_{N-2}b^2)^{5/8}=A_{N-2}b^2, \label{rooteq1}
\end{equation}
which we substitute into the other Bethe equations (\ref{bethe1.1})
and (\ref{bethe1.2}) in this limit, to give
\begin{equation}
\left[\frac{1}{\omega} (A_{N-2}b^2)^{5/8}\right]^2=\frac{( A_{N-2}b^2)^2}
{b^{2N}}\quad \Rightarrow \quad b^{2N}=1.
\end{equation}
Note that the value $b=-1$, determined numerically for the leading
2-string excitation, satisfies this equation. In principle these
equations can be solved consistently to determine the total number
of 2-string excitations.  

To proceed with the calculation we use the auxiliary functions 
$A(w)$, $X(w)$, $Y(1/w)$ defined in  (\ref{defA}), (\ref{defX}) and
(\ref{defY}) with the appropriate choices $s=5,\,r=16$, and define 
in analogy the ``unknown functions'',
\begin{eqnarray}
F_1(w)&=&\prod_{j=1}^{N-2}\prod_{k=0}^{\infty} (1-x^{32k}w/a_j), \nonumber \\
G_1(1/w)&=&\prod_{j=1}^{N-2}\prod_{k=1}^{\infty} (1-x^{32k}a_j/w), 
\label{def1a} \end{eqnarray}
along with the functions we will actually solve for,
\begin{eqnarray}
\F_1(w)&=&F_1(w)/F_1(x^{12}w), \nonumber\\
\G_1(1/w)&=&G_1(1/w)/G_1(1/x^{12}w).\label{def1b}
\end{eqnarray}
In fact, although we must define $F_i(w)$ and $G_i(w)$ for each $m_i$, we
can define 
\begin{eqnarray}
\F_i(w)&=&F_i(w)/F_i(x^{12}w), \nonumber\\
\G_i(1/w)&=&G_i(1/w)/G_i(1/x^{12}w), \label{bigfg}
\end{eqnarray}
for $i \neq 4,6$. Rearranging the Bethe equation (\ref{bethe1.k}) and 
equating the dominant terms, as in regime 1, gives 
\begin{eqnarray}
\F_1(a)&=&\frac{A(x^{10}a)}{A(x^{22}a)} 
\frac{X(x^{21}a)X(x^{23}a)}{X(x^9a)X(x^{11}a)}
\frac{\F_1(x^{10}a)}{\F_1(x^{20}a)}, \nonumber \\
&&\nonumber\\
\G_1(1/a)&=&\frac{A(x^{42}/a)}{A(x^{30}/a)} 
\frac{Y(1/x^{3}a)Y(1/xa)}{Y(x^9/a)Y(x^{11}/a)}
\frac{\G_1(x^{10}/a)}{\G_1(x^{20}/a)} .
\end{eqnarray}
We solve these recursively, using identities for $X(w)$ and $Y(1/w)$ 
given in Appendix A, so obtaining
\begin{eqnarray}
\F_1(a)&=&\F_0(a) \frac{X(x^{23}a)X(x^{33}a)}{X(x^{11}a)X(x^{53}a)}
\nonumber\\
&&\times\prod_{m=0}^{\infty}\left\{
\frac{(1-x^{60m+31}a/b)(1-x^{60m+39}a/b)}
{(1-x^{60m+9}a/b)(1-x^{60m+19}a/b)}\right.\nonumber\\
&&\phantom{\times\prod_{m=0}^{\infty}}\times \left.\frac{(1-x^{60m+49}a/b)
(1-x^{60m+81}a/b)}
{(1-x^{60m+51}a/b)(1-x^{60m+61}a/b)}\right\}, \label{f1}
\end{eqnarray}
and similarly
\begin{eqnarray}
\G_1(1/a)&=&\G_0(1/a)
\frac{Y(1/xa)Y(x^{41}/a)}{Y(x^{11}/a)Y(x^{21}/a)}
\nonumber \\
&&\times\prod_{m=1}^{\infty}\left\{
\frac{(1-x^{60m-31}b/a)(1-x^{60m-21}b/a)}
{(1-x^{60m-9}b/a)(1-x^{60m-1}b/a)}\right.\nonumber\\
&&\phantom{\times\prod_{m=1}^{\infty}}\times\left.
\frac{(1-x^{60m+11}b/a)(1-x^{60m+21}b/a)}{(1-x^{60m+9}b/a)(1-x^{60m+41}b/a)}
\right\}. \label{g1}
\end{eqnarray}
We are now in a position to evaluate the eigenvalue (\ref{BAEc}). 
Expressed in terms of the auxiliary functions and using (\ref{rooteq1}) it
reads 
\begin{eqnarray}
\lefteqn{\Lambda_1=\frac{w^2}{b^2}\frac{A(x^{2}w)A(x^{12}w)
A(x^{20}/w)A(x^{30}/w)}
{A(x^{2})A(x^{12})A(x^{20})A(x^{30})}} \nonumber \\
&&\phantom{+}\times \frac{X(x^{21}w)X(x^{31}w)}{X(xw)X(x^{11}w)}
\frac{Y(1/x^{31}w)Y(1/x^{21}w)}{Y(1/x^{11}w)Y(1/xw)}
\F_1(x^{10}w)\G_1(1/x^{10}w) \nonumber \\
&&+\frac{A(w)A(x^{2}w)A(x^{30}/w)A(x^{32}/w)}
{A(x^{2})A(x^{12})A(x^{20})A(x^{30})} 
\frac{X(w/x^{11})X(w/x^9)X(x^{11}w)X(x^{13}w)}{X(w/x^{21})X(xw)^2X(x^{23}w)}
\nonumber \\
&&\phantom{+}\times\frac{Y(1/x^{13}w)Y(1/x^{11}w)Y(x^9/w)Y(x^{11}/w)}
{Y(1/x^{23}w)Y(1/xw)^2Y(x^{21}/w)}
\frac{\F_1(w)\G_1(1/w)}{\F_1(w/x^{10})\G_1(x^{10}/w)} \nonumber \\
&&+\frac{A(w/x^{10})A(w)A(x^{32}/w)A(x^{42}/w)}
{A(x^{2})A(x^{12})A(x^{20})A(x^{30})}
\frac{X(w/x^{19})X(x^{3}w)}{X(w/x^{31})X(w/x^{9})} \nonumber \\
&&\phantom{+}\times
\frac{Y(1/x^{3}w)Y(x^{19}/w)}{Y(x^9/w)Y(x^{31}/w)}
\frac{1}{\F_1(w/x^{20})\G_1(x^{20}/w)}.
\end{eqnarray}
Using (\ref{f1}) and (\ref{g1}) we find that this result may be written 
in terms of elliptic functions as\footnote{We point out here that all
three terms in the eigenvalue expression
are identical, once $\F_1$ and $\G_1$ are substituted. This feature is
common to all masses $m_ i\,(i=1,\ldots, 8)$ when the appropriate functions
$\F_i$ and $\G_i$ have been found and used. So in discussion of the subsequent
masses, we suppress the second and third
terms of the eigenvalue expression.}
\begin{eqnarray}
\frac{\Lambda_1}{\Lambda_0} &=& \frac{w^2}{b^2}\,
\frac{E(xb/w,x^{60})\,E(x^{11}b/w,x^{60})}
     {E(xw/b,x^{60})\,E(x^{11}w/b,x^{60})}\nonumber\\
&&\phantom{w^2\,}\times\frac{E(x^{31}w/b,x^{60})\,E(x^{41}w/b,x^{60})}
{E(x^{31}b/w,x^{60})\,E(x^{41}b/w,x^{60})} .
\label{reg21b}
\end{eqnarray}
Setting $b=-1$, the leading excitation for $L=3$ in regime 2, which
clearly lies in the $w^2$ band, is thus
\begin{eqnarray}
\frac{\Lambda_1}{\Lambda_0} &=& w^2\,
\frac{E(-x/w,x^{60})\,E(-x^{11}/w,x^{60})}
     {E(-x\,w,x^{60})\,E(-x^{11}\,w,x^{60})}\nonumber\\
&&\phantom{w^2\,}\times\frac{E(-x^{31}\,w,x^{60})\,E(-x^{41}\,w,x^{60})}
{E(-x^{31}/w,x^{60})\,E(-x^{41}/w,x^{60})} .
\label{reg21}
\end{eqnarray}
At the isotropic point $w = x^{15}$ this reduces to
\begin{eqnarray}
\frac{\Lambda_1}{\Lambda_0} &=& x^{28}\,
\frac{E^2(-x^4,x^{60})\,E^2(-x^{14},x^{60})}
     {E^2(-x^{16},x^{60})\,E^2(-x^{26},x^{60})} \nonumber\\
&=& 
\left[
\frac{\tv(\frac{\pi}{15},p^{8/15})\,\tv(\frac{7\pi}{30},p^{8/15})}
     {\tv(\frac{4\pi}{15},p^{8/15})\,\tv(\frac{13\pi}{30},p^{8/15})} 
\right]^2, 
\end{eqnarray}
where we have written the excitation in terms of the original nome.

\subsection{Mass $m_2$}
We begin the perturbation argument with the structure
$w_j=a_j$ for $j=1, \ldots, N-4$ with $w_{N-3}=b_1x^{-5}$, 
$w_{N-2}=b_2x^{5}$, $w_{N-1}=b_3x^{-15}$ and
$w_N=b_4x^{15}$. From the Bethe equations for $k=N-3,\ldots,N$ 
(see Appendix B.2) we can show that $b_1=b_2=b_3=b_4=b$. 
The Bethe equation for the other roots $a_k=a$ is then
\begin{eqnarray}
\lefteqn{-\omega \left[a\frac{E(x^{10}/a)}{E(x^{10}a)}\right]^N=} 
\nonumber \\
&&(A_{N-4}b^4)^{5/8}\frac{a^4}{b^4}
\frac{E(x^3 b/a)E(x^5 b/a)}{E(x^3 a/b)E(x^5 a/b)} 
\prod_{j=1}^{N-4} \frac{E(x^{10}a/a_j)E(x^{20}a_j/a)} 
{E(x^{10}a_j/a)E(x^{20}a/a_j)}.\label{bethe2.k}
\end{eqnarray}
In the $x \to 0$ limit this gives the equation
\begin{equation}
a^{N-4}+\frac{1}{\omega} (A_{N-4}b^4)^{5/8}/b^4=0.
\end{equation}
Equating this as usual with
$\prod_{j=1}^{N-4}(a-a_j)$, we obtain
\begin{equation}
 \frac{1}{\omega} (A_{N-4}b^4)^{5/8}=A_{N-4}b^4,
\end{equation}
(which we later apply to prefactors in $\Lambda_2$).
From the other Bethe equations (\ref{bethe2.1})-(\ref{bethe2.4}) 
in this limit,
\begin{equation}
\left[\frac{1}{\omega} (A_{N-4}b^4)^{5/8}\right]^4=\frac{( A_{N-4}b^4)^4}
{b^{2N}}\quad \Rightarrow \quad b^{2N}=1.
\end{equation}

We define functions of the roots we wish to find as
\begin{eqnarray}
F_2(w)&=&\prod_{j=1}^{N-4}\prod_{k=0}^{\infty} (1-x^{32k}w/a_j),
\nonumber\\
G_2(1/w)&=&\prod_{j=1}^{N-4}\prod_{k=1}^{\infty} (1-x^{32k}a_j/w). 
\label{def2}
\end{eqnarray}
Treating the Bethe equation (\ref{bethe2.k}) as before gives, in terms 
of the functions defined in (\ref{bigfg}), the recurrences
\begin{eqnarray}
\F_2(a)&=&\frac{A(x^{10}a)}{A(x^{22}a)} 
\frac{X(x^{27}a)X(x^{29}a)}{X(x^{3}a)X(x^5a)}
\frac{\F_2(x^{10}a)}{\F_2(x^{20}a)}, \nonumber \\
&&\nonumber\\
\G_2(1/a)&=&\frac{A(x^{42}/a)}{A(x^{30}/a)} 
\frac{Y(1/x^{9}a)Y(1/x^{7}a)}{Y(x^{15}/a)Y(x^{17}/a)}
\frac{\G_2(x^{10}/a)}{\G_2(x^{20}/a)} .
\end{eqnarray}
Solving these we obtain
\begin{eqnarray}
\F_2(a)&=&\F_0(a) \frac{X(x^{29}a)X(x^{39}a)}{X(x^{5}a)X(x^{15}a)}
\nonumber \\
&&\times\prod_{m=0}^{\infty}\left\{
\frac{(1-x^{60m+27}a/b)(1-x^{60m+33}a/b)}
{(1-x^{60m+3}a/b)(1-x^{60m+13}a/b)}\right.\nonumber\\
&&\phantom{\times\prod_{m=0}^{\infty}}\times
\left.\frac{(1-x^{60m+37}a/b)(1-x^{60m+43}a/b)}
{(1-x^{60m+57}a/b)(1-x^{60m+67}a/b)}\right\},\label{f2}
\end{eqnarray}
and similarly
\begin{eqnarray}
\G_2(1/a)&=&\G_0(1/a)
\frac{Y(1/x^{7}a)Y(x^{3}/a)}{Y(x^{17}/a)Y(x^{27}/a)}
\nonumber \\
&&\times\prod_{m=1}^{\infty}\left\{
\frac{(1-x^{60m-37}b/a)(1-x^{60m-27}b/a)}
{(1-x^{60m-13}b/a)(1-x^{60m-7}b/a)}
\right.\nonumber\\
&&\phantom{\times\prod_{m=1}^{\infty}}\times\left.
\frac{(1-x^{60m+17}b/a)(1-x^{60m+27}b/a)}
{(1-x^{60m-3}b/a)(1-x^{60m+3}b/a)}\right\}.\label{g2}
\end{eqnarray}
We now substitute these into the eigenvalue expression, 
the first term of which gives 
\begin{eqnarray}
\Lambda_2&=&\frac{w^2}{b^2}\frac{A(x^{2}w)A(x^{12}w)A(x^{20}/w)A(x^{30}/w)}
{A(x^{2})A(x^{12})A(x^{20})A(x^{30})} 
\frac{X(x^{15}w)X(x^{25}w)}{X(x^{7}w)X(x^{17}w)}
\nonumber \\
&&\times
\frac{Y(1/x^{25}w)Y(1/x^{15}w)}{Y(1/x^{17}w)Y(1/x^{7}w)}
\F_2(x^{10}w)\G_2(1/x^{10}w).
\end{eqnarray}
This in turn gives an expression in elliptic functions analogous to
(\ref{reg21b}), but we immediately set $b=-1$, yielding the
leading 4-string excitation in the $w^2$ band for $L=3$ to be
\begin{eqnarray}
\frac{\Lambda_2}{\Lambda_0} &=& w^2\,
\frac{E(-x^{7}/w,x^{60})\,E(-x^{17}/w,x^{60})}
     {E(-x^{7}\,w,x^{60})\,E(-x^{17}\,w,x^{60})}
\nonumber\\
&&\phantom{w^2}\times\frac{E(-x^{37}\,w,x^{60})\,E(-x^{47}\,w,x^{60})}
{E(-x^{37}/w,x^{60})\,E(-x^{47}/w,x^{60})}.
\label{reg22}
\end{eqnarray}
At the isotropic point $w = x^{15}$ this reduces to
\begin{eqnarray}
\frac{\Lambda_2}{\Lambda_0} &=& x^{-10}\,
\frac{E^2(-x^2,x^{60})\,E^2(-x^{8},x^{60})}
     {E^2(-x^{22},x^{60})\,E^2(-x^{28},x^{60})} 
\nonumber\\
&=& 
\left[
\frac{\tv(\frac{\pi}{30},p^{8/15})\,\tv(\frac{2\pi}{15},p^{8/15})}
     {\tv(\frac{7\pi}{15},p^{8/15})\,\tv(\frac{11\pi}{30},p^{8/15})} 
\right]^2 .
\end{eqnarray}

\subsection{Mass $m_3$}
We begin the perturbation argument with the structure
$w_j=a_j$ for $j=1, \ldots, N-4$ with $w_{N-3}=b_1x^{-10}$, 
$w_{N-2}=b_2x^{10}$, $w_{N-1}=b_3x^{-20}$ and
$w_N=b_4x^{20}$. From the Bethe equations for $k=N-3,\ldots,N$
(see Appendix B.3) we can show that $b_1=b_2=b_3=b_4=b$. 
The Bethe equation for the other roots is
\begin{eqnarray}
\lefteqn{-\omega \left[a\frac{E(x^{10}/a)}{E(x^{10}a)}\right]^N=
(A_{N-4}b^4)^{5/8}\frac{a^4}{b^4}} \nonumber \\
&&\times\frac{E(x^8 b/a)E^2(x^{10}b/a)E(x^{12} b/a)}{E(x^8 a/b)E^2(x^{10}a/b)
E(x^{12} a/b)} 
\prod_{j=1}^{N-4} \frac{E(x^{10}a/a_j)E(x^{20}a_j/a)} 
{E(x^{10}a_j/a)E(x^{20}a/a_j)}.\label{bethe3.k}
\end{eqnarray}
In the $x \to 0$ limit this gives the equation
\begin{equation}
a^{N-4}+\frac{1}{\omega} (A_{N-4}b^4)^{5/8}/b^4=0,
\end{equation}
which we equate the lhs as usual with
$\prod_{j=1}^{N-4}(a-a_j)$ to obtain
\begin{equation}
 \frac{1}{\omega} (A_{N-4}b^4)^{5/8}=A_{N-4}b^4. 
\end{equation}
Using this with the other Bethe equations (\ref{bethe3.1})-(\ref{bethe3.4}) 
in the $x\rightarrow 0$ limit we obtain
\begin{equation}
\left[\frac{1}{\omega} (A_{N-4}b^4)^{5/8}\right]^4=\frac{( A_{N-4}b^4)^4}
{b^{3N}}\quad \Rightarrow \quad b^{3N}=1.
\end{equation}
Note that since $N$ is even, $b=-1$ satisfies this equation.

We define the appropriate functions for the unknown roots,
\begin{eqnarray}
F_3(w)&=&\prod_{j=1}^{N-4}\prod_{k=0}^{\infty} (1-x^{32k}w/a_j), \nonumber \\
G_3(1/w)&=&\prod_{j=1}^{N-4}\prod_{k=1}^{\infty} (1-x^{32k}a_j/w), \label{def3}
\end{eqnarray}
then from (\ref{bethe3.k}) come the recurrences,
\begin{eqnarray}
\F_3(a)&=&\frac{A(x^{10}a)}{A(x^{22}a)} 
\frac{X(x^{20}a)X^2(x^{22}a)X(x^{24}a)}{X(x^8a)X^2(x^{10}a)X(x^{12}a)}
\frac{\F_3(x^{10}a)}{\F_3(x^{20}a)}, \nonumber \\
&&\nonumber\\
\G_3(1/a)&=&\frac{A(x^{42}/a)}{A(x^{30}/a)} 
\frac{Y(1/x^{4}a)Y^2(1/x^2a)Y(1/a)}{Y(x^{8}/a)Y^2(x^{10}/a)Y(x^{12}/a)}
\frac{\G_3(x^{10}/a)}{\G_3(x^{20}/a)}.
\end{eqnarray}
The solutions are
\begin{eqnarray}
\F_3(a)&=&\F_0(a) \frac{X(x^{24}a)X(x^{32}a)X(x^{34}a)}
{X(x^{10}a)X(x^{12}a)X(x^{20}a)}
\nonumber \\
&&\times\prod_{m=0}^{\infty}\left\{
\frac{(1-x^{60m+22}a/b)(1-x^{60m+30}a/b)(1-x^{60m+32}a/b)}
{(1-x^{60m+8}a/b)(1-x^{60m+10}a/b)(1-x^{60m+18}a/b)}\right.\nonumber\\
&&\times
\left.\frac{(1-x^{60m+38}a/b)(1-x^{60m+40}a/b)(1-x^{60m+48}a/b)}
{(1-x^{60m+52}a/b)(1-x^{60m+60}a/b)(1-x^{60m+62}a/b)}\right\},\label{f3}
\end{eqnarray}
\begin{eqnarray}
\G_3(1/a)&=&\G_0(1/a)
\frac{Y(1/a)Y(1/x^2a)Y(x^8/a)}{Y(x^{12}/a)Y(x^{20}/a)Y(x^{22}/a)}
\nonumber \\
&&\times\prod_{m=1}^{\infty}\left\{
\frac{(1-x^{60m-32}b/a)(1-x^{60m-30}b/a)(1-x^{60m-22}b/a)}
{(1-x^{60m-18}b/a)(1-x^{60m-10}b/a)(1-x^{60m-8}b/a)}
\right.\nonumber\\
&&\times\left.
\frac{(1-x^{60m+12}b/a)(1-x^{60m+20}b/a)(1-x^{60m+22}b/a)}
{(1-x^{60m-2}b/a)(1-x^{60m}b/a)(1-x^{60m+8}b/a)}\right\}. \label{g3}
\end{eqnarray}
In terms of these functions the eigenvalue may be represented as
\begin{eqnarray}
\Lambda_3&=&-\frac{w^3}{b^3}\frac{A(x^{2}w)A(x^{12}w)A(x^{20}/w)
A(x^{30}/w)}{A(x^{2})A(x^{12})A(x^{20})A(x^{30})}
\frac{X(x^{20}w)X(x^{22}w)X(x^{30}w)}{X(x^{2}w)X(x^{10}w)X(x^{12}w)} 
\nonumber \\
&&
\times\frac{Y(1/x^{30}w)Y(1/x^{22}w)Y(1/x^{20}w)}
{Y(1/x^{12}w)Y(1/x^{10}w)Y(1/x^{2}w)}
\F_3(x^{10}w)\G_3(1/x^{10}w). 
\end{eqnarray}
Thus, application of the perturbation argument has yielded the
leading excitation in the $w^3$ band to be
\begin{eqnarray}
\frac{\Lambda_3}{\Lambda_0} &=& w^3\,
\frac{E(-x^{2}/w,x^{60})\,E(-x^{10}/w,x^{60})\,E(-x^{12}/w,x^{60})}
     {E(-x^{2}\,w,x^{60})\,E(-x^{10}\,w,x^{60})\,E(-x^{12}\,w,x^{60})}
\nonumber\\
&&\times\frac{E(-x^{32}\,w,x^{60})\,E(-x^{40}\,w,x^{60})
E(-x^{42}\,w,x^{60})}
{E(-x^{32}/w,x^{60})\,E(-x^{40}/w,x^{60})E(-x^{42}/w,x^{60})},
\label{reg23}
\end{eqnarray}
where we have put $b=-1$. At the isotropic point $w = x^{15}$ this reduces to
\begin{equation}
\frac{\Lambda_3}{\Lambda_0} =
\left[
\frac{\tv(\frac{13\pi}{60},p^{8/15})\,\tv(\frac{\pi}{12},p^{8/15})
\,\tv(\frac{\pi}{20},p^{8/15})}
     {\tv(\frac{17\pi}{60},p^{8/15})\,\tv(\frac{5\pi}{12},p^{8/15})
\,\tv(\frac{9\pi}{20},p^{8/15})} 
\right]^2 .
\end{equation}

\subsection{Mass $m_4$}
We begin the perturbation argument with the 7-string 
$w_j=a_j$ for $j=1, \ldots, N-7$ with 
$w_{N-6}=b_1x^{-1}$, $w_{N-5}=b_2x$, $w_{N-4}=b_3x^{-7}$,
$w_{N-3}=b_4x^{7}$, $w_{N-2}=b_5x^{-13}$, $w_{N-1}=b_6x^{13}$ and
$w_N=b_7x^{16}$. From the Bethe equations for $k=N-6,\ldots,N$,
we show in Appendix B.4 that $b_1=b_4=b_5=\alpha$ and $b_2=b_3=b_6=
\beta$. We further let $b_7=b$. This feature is different
from that seen for the previous masses (or will
see for $m_5,\,m_7,\,m_8$). However, in the final analysis, our
answer does not depend upon $\alpha$ or $\beta$.

The other Bethe roots satisfy
\begin{eqnarray}
\lefteqn{-\omega \left[a\frac{E(x^{10}/a)}{E(x^{10}a)}\right]^N=
(A_{N-7}\alpha^3 \beta^3 b)^{5/8}\frac{a^6}{\alpha^2 \beta^2 b^2}} 
\nonumber \\ &&\times
\frac{E(x^{7}\alpha/a)E(x^{21} \alpha/a)E(x^{27}\alpha/a)}
{E(xa/\alpha)E(x^{15} a/\alpha)E(x^{21}a/\alpha)}
\frac{E(x\beta/a)E(x^{15} \beta/a)E(x^{21}\beta/a)} 
{E(x^{7}a/\beta)E(x^{21} a/\beta)E(x^{27}a/\beta)} \nonumber \\
&&\times \frac{E(x^4 b/a)E(x^{6} b/a)} 
{E(x^{4}a/b)E(x^{6} a/b)}
 \prod_{j=1}^{N-7} \frac{E(x^{10}a/a_j)E(x^{20}a_j/a)} 
{E(x^{10}a_j/a)E(x^{20}a/a_j)}. \label{bethe4.k}
\end{eqnarray}
In the $x \to 0$ limit this gives the equation
\begin{equation}
a^{N-6}+\frac{1}{\omega} (A_{N-7}\alpha^3 \beta^3 b)^{5/8}/\alpha^2 
\beta^2 b^2=0, \label{a4eq}
\end{equation}
which is an $(N-6)\th$ order equation for $N-7$ zeros, so that
there will be a hole, $a_{N-6}$, in this case, as there was
in regime 1. Equating the lhs of (\ref{a4eq}) with
$\prod_{j=1}^{N-6}(a-a_j)$ we obtain
\begin{equation}
 \frac{1}{\omega} (A_{N-7}\alpha^3 \beta^3 b)^{5/8}=A_{N-6}\alpha^2 
\beta^2 b^2 =A_{N-7}a_{N-6}\alpha^2 \beta^2 b^2. \label{aseq}
\end{equation}
From the other Bethe equations (\ref{bethe4.1})-(\ref{bethe4.7}) 
in the $x \rightarrow 0$ limit,
\begin{equation}
\left(-\frac{1}{\omega} (A_{N-7}\alpha^3 \beta^3 b)^{5/8}\right)^7=
\frac{(A_{N-7})^6(\alpha^2\beta^2b^2)^7}{b^{2N}}.
\label{beq}
\end{equation}
Now (\ref{a4eq}) must be satisfied by each of the 
$a_j$, $j=1, \ldots, N-7$,
\begin{equation}
a_j^{N-6}=-A_{N-7}a_{N-6},
\end{equation}
so that forming the product of these $(N-7)$ equations,
\begin{equation}
(A_{N-7})^{N-6}=-(A_{N-7}a_{N-6})^{N-7} \Rightarrow (a_{N-6})^{N}=
-A_{N-7}(a_{N-6})^7.\label{holeeq}
\end{equation}
Taking  (\ref{aseq}), (\ref{beq}) and (\ref{holeeq}) together gives 
\begin{equation}
(b^2a_{N-6})^N=1,
\end{equation}
which is clearly satisfied by $b=-1$ and $a_{N-6}=-1$.
We need to define, in analogy with the previous auxiliary functions,
\begin{eqnarray}
X_\alpha(w)&=&\prod_{k=0}^{\infty}(1-x^{32k}w/\alpha), \quad
Y_\alpha(1/w)=\prod_{k=1}^{\infty}(1-x^{32k}\alpha/w), \label{xalpha}\\
X_\beta(w)&=&\prod_{k=0}^{\infty}(1-x^{32k}w/\beta), 
\quad Y_\beta(1/w)=\prod_{k=1}^{\infty}(1-x^{32k}\beta/w), \\
R_4(w)&=&\prod_{k=0}^{\infty}(1-x^{32k}w/a_{N-6}), \\
S_4(1/w)&=&\prod_{k=1}^{\infty}(1-x^{32k}a_{N-6}/w),
\end{eqnarray}
and the functions 
\begin{eqnarray}
F_4(w)&=&\prod_{j=1}^{N-6}\prod_{k=0}^{\infty} (1-x^{32k}w/a_j), \nonumber \\
G_4(1/w)&=&\prod_{j=1}^{N-6}\prod_{k=1}^{\infty} (1-x^{32k}a_j/w),
\end{eqnarray}
for the as-yet-unknown roots. In place of the usual definition, we have 
\begin{eqnarray}
\F_4(w)&=&\frac{F_4(w)}{F_4(x^{12}w)}
\frac{X_\alpha(xw)X_\beta(x^7w)}{X_\alpha(x^5w)X_\beta(x^{11}w)},\nonumber\\
\G_4(w)&=&\frac{G_4(1/w)}{G_4(1/x^{12}w)}
\frac{Y_\alpha(1/xw)Y_\beta(1/x^7w)}{Y_\alpha(1/x^5w)Y_\beta(1/x^{11}w)}.
\label{bigfg4} \end{eqnarray}
Rearranging the Bethe equations (\ref{bethe4.k}) gives
recurrences as usual, with
\begin{eqnarray}
\F_4(a)&=&\frac{A(x^{10}a)}{A(x^{22}a)} 
\frac{X(x^{26}a)X(x^{28}a)R_4(x^{20}a)R_4(x^{22}a)}
{X(x^{4}a)X(x^{6}a)R_4(x^{10}a)R_4(x^{12}a)}
\frac{\F_4(x^{10}a)}{\F_4(x^{20}a)}, \nonumber\\
&&\nonumber\\
\G_4(1/a)&=&\frac{A(x^{42}/a)}{A(x^{30}/a)} 
\frac{Y(1/x^{6}a)Y(1/x^8a)S_4(1/x^{2}a)S_4(1/a)}
{Y(x^{14}/a)Y(x^{16}/a)S_4(x^{8}/a)S_4(x^{10}/a)}
\frac{\G_4(x^{10}/a)}{\G_4(x^{20}/a)}, 
\end{eqnarray}
which we solve using identities for the various auxiliary functions
listed in Appendix A to obtain
\begin{eqnarray}
\F_4(a)&=&\F_0(a) \frac{X(x^{28}a)X(x^{38}a)}
{X(x^{6}a)X(x^{16}a)}\frac{R_4(x^{32}a)}{R_4(x^{12}a)}
\nonumber \\
&&\times\prod_{m=0}^{\infty}\left\{
\frac{(1-x^{60m+26}\frac{a}{b})(1-x^{60m+34}\frac{a}{b})
(1-x^{60m+36}\frac{a}{b})}
{(1-x^{60m+4}\frac{a}{b})(1-x^{60m+14}\frac{a}{b})
(1-x^{60m+56}\frac{a}{b})}\right.\nonumber\\
&&\times
\left.\frac{(1-x^{60m+44}\frac{a}{b})}{(1-x^{60m+66}\frac{a}{b})}
\frac{(1-x^{60m+30}\frac{a}{a_{N-6}})(1-x^{60m+40}\frac{a}{a_{N-6}})}
{(1-x^{60m+10}\frac{a}{a_{N-6}})(1-x^{60m+60}\frac{a}{a_{N-6}})}\right\},
\end{eqnarray} \label{f4}
and similarly
\begin{eqnarray}
\G_4(1/a)&=&\G_0(1/a)
\frac{Y(1/x^6a)Y(x^4/a)}{Y(x^{16}/a)Y(x^{26}/a)}
\frac{S_4(1/a)}{S_4(x^{20}/a)}
\nonumber \\
&&\times\prod_{m=1}^{\infty}\left\{
\frac{(1-x^{60m-36}\frac{b}{a})(1-x^{60m-26}\frac{b}{a})(1-x^{60m+16}
\frac{b}{a})}
{(1-x^{60m-14}\frac{b}{a})(1-x^{60m-6}\frac{b}{a})(1-x^{60m-4}\frac{b}{a})}
\right.\nonumber\\
&&\times\left.
\frac{(1-x^{60m+26}\frac{b}{a})(1-x^{60m-30}\frac{a_{N-6}}{a})
(1-x^{60m+20}\frac{a_{N-6}}{a})}
{(1-x^{60m+4}\frac{b}{a})(1-x^{60m-10}\frac{a_{N-6}}{a})
(1-x^{60m}\frac{a_{N-6}}{a})}\right\}. \label{g4}
\end{eqnarray}
The eigenvalue expression is
\begin{eqnarray}
\Lambda_4&=&-\frac{w^3}{a_{N-6}b^2}\frac{A(x^{2}w)A(x^{12}w)
A(x^{20}/w)A(x^{30}/w)}
{A(x^{2})A(x^{12})A(x^{20})A(x^{30})}\nonumber \\
&&\times\frac{X(x^{26}w)}{X(x^{6}w)}\frac{Y(1/x^{26}w)}{Y(1/x^{6}w)}
\F_4(x^{10}w)\G_4(1/x^{10}w), 
\end{eqnarray}
which, in elliptic functions is
\begin{eqnarray}
\frac{\Lambda_4}{\Lambda_0} &=& -\frac{w^3}{a_{N-6}b^2}\,
\frac{E(x^{6}b/w,x^{60})\,E(x^{10}a_{N-6}/w,x^{60})\,E(x^{14}b/w,x^{60})}
     {E(x^{6}w/b,x^{60})\,E(x^{10}w/a_{N-6},x^{60})\,E(x^{14}w/b,x^{60})}
\nonumber\\
&&\times\frac{E(x^{36}w/b,x^{60})\,E(x^{40}w/a_{N-6},x^{60})
E(x^{44}w/b,x^{60})}
{E(x^{36}b/w,x^{60})\,E(x^{40}a_{N-6}/w,x^{60})E(x^{44}b/w,x^{60})}. 
\label{reg24b}
\end{eqnarray}
Finally, with $b=a_{N-6}=-1$,
\begin{eqnarray}
\frac{\Lambda_4}{\Lambda_0} &=&w^3\,
\frac{E(-x^{6}/w,x^{60})\,E(-x^{10}/w,x^{60})\,E(-x^{14}/w,x^{60})}
     {E(-x^{6}\,w,x^{60})\,E(-x^{10}\,w,x^{60})\,E(-x^{14}\,w,x^{60})}
\nonumber\\
&&\times\frac{E(-x^{36}\,w,x^{60})\,E(-x^{40}\,w,x^{60})
E(-x^{44}\,w,x^{60})}
{E(-x^{36}/w,x^{60})\,E(-x^{40}/w,x^{60})E(-x^{44}/w,x^{60})} .
\label{reg24}
\end{eqnarray}
At the isotropic point $w = x^{15}$ this becomes
\begin{equation}
\frac{\Lambda_4}{\Lambda_0} =
\left[
\frac{\tv(\frac{3\pi}{20},p^{8/15})\,\tv(\frac{\pi}{12},p^{8/15})
\,\tv(\frac{\pi}{60},p^{8/15})}
     {\tv(\frac{7\pi}{20},p^{8/15})\,\tv(\frac{5\pi}{12},p^{8/15})
\,\tv(\frac{29\pi}{60},p^{8/15})} 
\right]^2 .
\end{equation}

\subsection{Mass $m_5$}
We begin the perturbation argument with 
$w_j=a_j$ for $j=1, \ldots, N-6$ and $w_{N-5}=b_1x^{-13}$, 
$w_{N-4}=b_2x^{13}$, $w_{N-3}=b_3x^{-11}$, $w_{N-2}=b_4x^{11}$,
$w_{N-1}=b_5x^{-9}$, $w_N=b_6x^{9}$. In Appendix B.5  
we show that the $b_i$ are equal, and we call them $b$. 
The Bethe equation for the other roots is
\begin{eqnarray}
\lefteqn{-\omega \left[a\frac{E(x^{10}/a)}{E(x^{10}a)}\right]^N=
(A_{N-6}b^6)^{5/8}\frac{a^6}{b^6}
\frac{E(x^{7}b/a)E^2(x^{9}b/a)}{E(x^7a/b)E^2(x^9 a/b)}} \nonumber \\
&&
\times\frac{E^2(x^{11} b/a)E(x^{13}b/a)}
{E^2(x^{11}a/b)E(x^{13}a/b)} 
\prod_{j=1}^{N-6} \frac{E(x^{10}a/a_j)E(x^{20}a_j/a)} 
{E(x^{10}a_j/a)E(x^{20}a/a_j)}. \label{bethe5.k}
\end{eqnarray}
In the $x \to 0$ limit this gives the equation
\begin{equation}
a^{N-6}+\frac{1}{\omega} (A_{N-6}b^6)^{5/8}/b^6=0,
\end{equation}
which is an $(N-6)\th$ order equation for $N-6$ zeros,
so that no hole needs to be considered. Equating the lhs
with $\prod_{j=1}^{N-6}(a-a_j)$ gives
\begin{equation}
 \frac{1}{\omega} (A_{N-6}b^6)^{5/8}=A_{N-6}b^6. 
\end{equation}
From the other Bethe equations (\ref{bethe5}),
\begin{equation}
\left[\frac{1}{\omega} (A_{N-6}b^6)^{5/8}\right]^6=\frac{( A_{N-6}b^6)^6}
{b^{4N}}\quad \Rightarrow \quad b^{4N}=1.
\end{equation}

For the unknown roots we define the functions
\begin{eqnarray}
F_5(w)&=&\prod_{j=1}^{N-6}\prod_{k=0}^{\infty} (1-x^{32k}w/a_j), 
\nonumber \\
G_5(1/w)&=&\prod_{j=1}^{N-6}\prod_{k=1}^{\infty} (1-x^{32k}a_j/w), 
\label{def5}
\end{eqnarray}
which, after rearranging (\ref{bethe5.k}), obey the recurrences
\begin{eqnarray}
\F_5(a)&=&\frac{A(x^{10}a)}{A(x^{22}a)} 
\frac{X(x^{19}a)X^2(x^{21}a)X^2(x^{23}a)X(x^{25}a)}
{X(x^7a)X^2(x^{9}a)X^2(x^{11}a)X(x^{13}a)}
\frac{\F_5(x^{10}a)}{\F_5(x^{20}a)}, \nonumber \\
&&\nonumber\\
\G_5(1/a)&=&\frac{A(x^{42}/a)}{A(x^{30}/a)} 
\frac{Y(1/x^5a)Y^2(1/x^3a)Y^2(1/xa)Y(x/a)}
{Y(x^{7}/a)Y^2(x^{9}/a)Y^2(x^{11}/a)Y(x^{13}/a)}
\frac{\G_5(x^{10}/a)}{\G_5(x^{20}/a)}. 
\end{eqnarray}
The solutions are
\begin{eqnarray}
\F_5(a)&=&\F_0(a) \frac{X(x^{25}a)X(x^{31}a)X(x^{33}a)X(x^{35}a)}
{X(x^{9}a)X(x^{11}a)X(x^{13}a)X(x^{19}a)}
\nonumber \\
&&\times\prod_{m=0}^{\infty}\left\{
\frac{(1-x^{60m+23}a/b)(1-x^{60m+29}a/b)(1-x^{60m+31}a/b)}
{(1-x^{60m+7}a/b)(1-x^{60m+9}a/b)(1-x^{60m+11}a/b)}
\right.\nonumber\\
&&\times
\left.\frac{(1-x^{60m+33}a/b)(1-x^{60m+37}a/b)(1-x^{60m+39}a/b)}
{(1-x^{60m+17}a/b)(1-x^{60m+53}a/b)(1-x^{60m+59}a/b)}\right.
\nonumber \\
&&\times
 \left.\frac{(1-x^{60m+41}a/b)(1-x^{60m+47}a/b)}
{(1-x^{60m+61}a/b)(1-x^{60m+63}a/b)}\right\}, \label{f5}
\end{eqnarray}
\begin{eqnarray}
\G_5(1/a)&=&\G_0(1/a)
\frac{Y(1/x^3a)Y(1/xa)Y(x/a)Y(x^7/a)}{Y(x^{13}/a)
Y(x^{19}/a)Y(x^{21}/a)Y(x^{23}/a)}
\nonumber \\
&&\times\prod_{m=1}^{\infty}\left\{
\frac{(1-x^{60m-33}b/a)(1-x^{60m-31}b/a)(1-x^{60m-29}b/a)}
{(1-x^{60m-17}b/a)(1-x^{60m-11}b/a)(1-x^{60m-9}b/a)}
\right.\nonumber\\
&&\times\left.
\frac{(1-x^{60m-23}b/a)(1-x^{60m+13}b/a)(1-x^{60m+19}b/a)}
{(1-x^{60m-7}b/a)(1-x^{60m-3}b/a)(1-x^{60m-1}b/a)}
\right.\nonumber \\
&&\times \left. \frac{(1-x^{60m+21}b/a)(1-x^{60m+23}b/a)}
{(1-x^{60m+1}b/a)(1-x^{60m+7}b/a)}\right\}, \label{g5}
\end{eqnarray}
which we next substitute into the eigenvalue expression
\begin{eqnarray}
\Lambda_5&=&\frac{w^4}{b^4}
            \frac{A(x^{2}w)A(x^{12}w)A(x^{20}/w)A(x^{30}/w)}
                 {A(x^{2})A(x^{12})A(x^{20})A(x^{30})}\nonumber\\
&&
\times\frac{X(x^{19}w)X(x^{21}w)X(x^{23}w)X(x^{29}w)}
{X(x^{3}w)X(x^{9}w)X(x^{11}w)X(x^{13}w)} \nonumber \\
&&
\times\frac{Y(1/x^{29}w)Y(1/x^{23}w)Y(1/x^{21}w)Y(1/x^{19}w)}
{Y(1/x^{3}w)Y(1/x^{9}w)Y(1/x^{11}w)Y(1/x^{13}w)} \nonumber \\
&&
\times \F_5(x^{10}w)\G_5(1/x^{10}w), 
\end{eqnarray}
to obtain (with $b=-1$)
\begin{eqnarray}
\frac{\Lambda_5}{\Lambda_0} &=& w^4
\frac{E(-x^{3}/w)E(-x^{9}/w)E(-x^{11}/w)E(-x^{13}/w)}
     {E(-x^{3}w)E(-x^{9}w)E(-x^{11}w)E(-x^{13}w)}
\nonumber\\
&&\times
\frac{E(-x^{33}w)E(-x^{39}w)E(-x^{41}w)E(-x^{43}w)}
{E(-x^{33}/w)E(-x^{39}/w)E(-x^{41}/w)E(-x^{43}/w)}. \label{reg25}
\end{eqnarray}
The elliptic functions are of nome $x^{60}$.
At the isotropic point $w = x^{15}$ the leading excitation reduces to
\begin{equation}
\frac{\Lambda_5}{\Lambda_0} =
\left[
\frac{\tv(\frac{\pi}{5},p^{8/15})\,\tv(\frac{\pi}{10},p^{8/15})
\,\tv(\frac{\pi}{15},p^{8/15})\,\tv(\frac{\pi}{30},p^{8/15})}
     {\tv(\frac{3\pi}{10},p^{8/15})\,\tv(\frac{2\pi}{5},p^{8/15})
\,\tv(\frac{13\pi}{30},p^{8/15})\,\tv(\frac{7\pi}{15},p^{8/15})} 
\right]^2 .
\end{equation}

\subsection{Mass $m_6$}
We begin with 
$w_j=a_j$ for $j=1, \ldots, N-7$ and
$w_{N-6}=b_1x^{-14}$, $w_{N-5}=b_2x^{14}$, $w_{N-4}=b_3x^{-10}$,
$w_{N-3}=b_4x^{10}$, $w_{N-2}=b_5x^{-6}$, $w_{N-1}=b_6x^{6}$,
$w_N=b_7x^{16}$. In Appendix B.6 we show that $b_1=b_2=b_5=b_6=b_7=b$ 
and that $b_3=b_4=\alpha$. This feature is similar to that for
$m_4$; again, in the final analysis, our
answer does not depend  upon $\alpha$. The Bethe equation for the 
other roots is
\begin{eqnarray}
\lefteqn{-\omega \left[a\frac{E(x^{10}/a)}{E(x^{10}a)}\right]^N=
(A_{N-7}b^5 \alpha^2)^{5/8}\frac{a^7}{b^6 \alpha}} \nonumber \\
&&\times
\frac{E(x^{2}b/a)E(x^{4} b/a)E(x^{6}b/a)E(x^{8}b/a)
E(x^{14}b/a)}
{E(x^2a/b)E(x^{4} a/b)E(x^{6}a/b)E(x^{8}a/b)E(x^{14}a/b)}
 \nonumber \\
&&\times \frac{E(x^{10}\alpha/a)E(x^{12} \alpha/a)E(x^{30}\alpha/a)} 
{E(x^{10}a/\alpha)E(x^{12} a/\alpha)E(x^{30}a/\alpha)}
 \prod_{j=1}^{N-7} \frac{E(x^{10}a/a_j)E(x^{20}a_j/a)} 
{E(x^{10}a_j/a)E(x^{20}a/a_j)}. \label{bethe6.k}
\end{eqnarray}
In the $x \to 0$ limit this gives 
\begin{equation}
a^{N-7}+\frac{1}{\omega} (A_{N-7}b^5 \alpha^2)^{5/8}/b^6\alpha=0
\end{equation}
as the equation for the $(N-7)$ roots.
Equating this with $\prod_{j=1}^{N-7}(a-a_j)$ gives
\begin{equation}
 \frac{1}{\omega} (A_{N-7}b^5 \alpha^2)^{5/8}=A_{N-7}b^6 \alpha.
\end{equation}
From the other Bethe equations (\ref{bethe6}) we obtain 
\begin{equation}
\left[\frac{1}{\omega} (A_{N-7}b^5\alpha^2)^{5/8}\right]^7=
\frac{( A_{N-7}b^6\alpha)^7}
{b^{4N}}\quad \Rightarrow \quad b^{4N}=1.
\end{equation}
We define 
\begin{eqnarray}
F_6(w)&=&\prod_{j=1}^{N-7}\prod_{k=0}^{\infty} (1-x^{32k}w/a_j), 
\nonumber \\
G_6(1/w)&=&\prod_{j=1}^{N-7}\prod_{k=1}^{\infty} (1-x^{32k}a_j/w), 
\label{def6a}
\end{eqnarray}
and in place of the usual definition, 
\begin{eqnarray}
\F_6(w)&=&\frac{F_6(w)}{F_6(x^{12}w)}
\frac{X_\alpha(x^{10}w)}{X_\alpha(x^{2}w)},\nonumber \\
\G_6(w)&=&\frac{G_6(1/w)}{G_6(1/x^{12}w)}
\frac{Y_\alpha(1/x^{10}w)}{Y_\alpha(1/x^{2}w)}, \label{def6b}
\end{eqnarray}
where $X_\alpha(w)$ and $Y_\alpha(1/w)$ have the same definition 
(\ref{xalpha}) as they did for $m_4$.
We solve the recurrences
\begin{eqnarray}
\F_6(a)&=&\frac{A(x^{10}a)}{A(x^{22}a)} 
\frac{X(x^{18}a)X(x^{24}a)X(x^{26}a)X(x^{28}a)X(x^{30}a)}
{X(x^{2}a)X(x^{4}a)X(x^{6}a)X(x^{8}a)X(x^{14}a)}
\frac{\F_6(x^{10}a)}{\F_6(x^{20}a)}, \nonumber \\
&&\nonumber\\
\G_6(1/a)&=&\frac{A(x^{42}/a)}{A(x^{30}/a)} 
\frac{Y(1/x^{10}a)Y(1/x^8a)Y(1/x^6a)Y(1/x^4a)Y(x^2/a)}
{Y(x^{6}/a)Y(x^{12}/a)Y(x^{14}/a)Y(x^{16}/a)Y(x^{18}/a)}
 \nonumber\\
&& \times 
\frac{\G_6(x^{10}/a)}{\G_6(x^{20}/a)},
\end{eqnarray}
which are obtained from (\ref{bethe6.k}), giving
\begin{eqnarray}
\F_6(a)&=&\F_0(a) \frac{X(x^{28}a)X(x^{30}a)}
{X(x^{14}a)X(x^{16}a)}
\nonumber \\
&&\times\prod_{m=0}^{\infty}\left\{
\frac{(1-x^{60m+26}\frac{a}{b})(1-x^{60m+28}\frac{a}{b})
(1-x^{60m+32}\frac{a}{b})}
{(1-x^{60m+2}\frac{a}{b})(1-x^{60m+4}\frac{a}{b})
(1-x^{60m+6}\frac{a}{b})}\right.\nonumber\\
&&\times
\left.\frac{(1-x^{60m+34}\frac{a}{b})(1-x^{60m+36}\frac{a}{b})
(1-x^{60m+38}\frac{a}{b})}
{(1-x^{60m+8}\frac{a}{b})(1-x^{60m+12}\frac{a}{b})(1-x^{60m+14}\frac{a}{b})}
\right. \nonumber \\
&& \times \left.\frac{(1-x^{60m+42}\frac{a}{b})(1-x^{60m+44}\frac{a}{b})}
{(1-x^{60m+56}\frac{a}{b})(1-x^{60m+58}\frac{a}{b})}\right\},
\label{f6}
\end{eqnarray}
\begin{eqnarray}
\G_6(1/a)&=&\G_0(1/a)
\frac{Y(x^2/a)Y(x^4/a)}{Y(x^{16}/a)Y(x^{18}/a)}
\nonumber \\
&&\times\prod_{m=1}^{\infty}\left\{
\frac{(1-x^{60m-38}\frac{b}{a})(1-x^{60m-36}\frac{b}{a})
(1-x^{60m-34}\frac{b}{a})}
{(1-x^{60m-14}\frac{b}{a})(1-x^{60m-12}\frac{b}{a})(1-x^{60m-8}\frac{b}{a})}
\right.\nonumber\\
&&\times\left.
\frac{(1-x^{60m-32}\frac{b}{a})(1-x^{60m-28}\frac{b}{a})
(1-x^{60m-26}\frac{b}{a})}
{(1-x^{60m-6}\frac{b}{a})(1-x^{60m-4}\frac{b}{a})(1-x^{60m-2}\frac{b}{a})}
\right. \nonumber \\
&& \times \left.\frac{(1-x^{60m+16}\frac{b}{a})(1-x^{60m+18}\frac{b}{a})}
{(1-x^{60m+2}\frac{b}{a})(1-x^{60m+4}\frac{b}{a})} \right\}.
\label{g6}
\end{eqnarray}
Substituted into the eigenvalue expression
\begin{eqnarray}
\Lambda_6&=&\frac{w^4}{b^4}\frac{A(x^{2}w)A(x^{12}w)A(x^{20}/w)A(x^{30}/w)}
{A(x^{2})A(x^{12})A(x^{20})A(x^{30})}\nonumber \\
&&\times\frac{X(x^{24}w)X(x^{26}w)}{X(x^{6}w)X(x^{8}w)}
\frac{Y(1/x^{26}w)Y(1/x^{24}w)}{Y(1/x^{8}w)Y(1/x^{6}w)} \nonumber \\
&&\times
\F_6(x^{10}w)\G_6(1/x^{10}w), 
\end{eqnarray}
this gives (with $b=-1$ and with elliptic nome $x^{60}$),
\begin{eqnarray}
\frac{\Lambda_6}{\Lambda_0} &=& w^4
\frac{E(-x^{6}/w)E(-x^{8}/w)E(-x^{12}/w)E(-x^{14}/w)}
     {E(-x^{6}w),E(-x^{8}w),E(-x^{12}w)E(-x^{14}w)}
\nonumber\\
&&\times\frac{E(-x^{36}w)E(-x^{38}w)E(-x^{42}w)E(-x^{44}w)}
{E(-x^{36}/w)E(-x^{38}/w)E(-x^{42}/w)E(-x^{44}/w)}. 
\label{reg26}
\end{eqnarray}
At the isotropic point this reduces to
\begin{equation}
\frac{\Lambda_6}{\Lambda_0} =
\left[
\frac{\tv(\frac{3\pi}{20},p^{8/15})\,\tv(\frac{7\pi}{60},p^{8/15})
\,\tv(\frac{\pi}{20},p^{8/15})\,\tv(\frac{\pi}{60},p^{8/15})}
     {\tv(\frac{7\pi}{20},p^{8/15})\,\tv(\frac{23\pi}{60},p^{8/15})
\,\tv(\frac{9\pi}{20},p^{8/15})\,\tv(\frac{29\pi}{60},p^{8/15})} 
\right]^2 .
\end{equation}

\subsection{Mass $m_7$}
We begin with 
$w_j=a_j$ for $j=1, \ldots, N-8$ and $w_{N-7}=b_1x^{-14}$, 
$w_{N-6}=b_2x^{14}$, $w_{N-5}=b_3x^{-12}$, $w_{N-4}=b_4x^{12}$,
$w_{N-3}=b_5x^{-10}$, $w_{N-2}=b_6x^{10}$, $w_{N-1}=b_7x^{-8}$,
$w_N=b_8x^{8}$. We show in Appendix B.7 that the $b_i$
are all equal. The Bethe equation for the other roots is
\begin{eqnarray}
\lefteqn{-\omega \left[a\frac{E(x^{10}/a)}{E(x^{10}a)}\right]^N=
(A_{N-8}b^8)^{5/8}\frac{a^8}{b^8}
\frac{E(x^{6}b/a)E^2(x^{8}b/a)E^2(x^{10}b/a)}{E(x^6a/b)E^2(x^8 a/b)
E^2(x^{10} a/b)}} \nonumber \\
&&
\times\frac{E^2(x^{12} b/a)E(x^{14}b/a)}
{E^2(x^{12}a/b)E(x^{14}a/b)} 
\prod_{j=1}^{N-8} \frac{E(x^{10}a/a_j)E(x^{20}a_j/a)} 
{E(x^{10}a_j/a)E(x^{20}a/a_j)}. \label{bethe7.k}
\end{eqnarray}
In the $x \to 0$ limit this gives 
\begin{equation}
a^{N-8}+\frac{1}{\omega} (A_{N-8}b^8)^{5/8}/b^8=0,
\end{equation}
which is an $(N-8)\th$ order equation for $N-8$ zeros, so that
there is again no hole. Equating this with
$\prod_{j=1}^{N-8}(a-a_j)$ we obtain
\begin{equation}
 \frac{1}{\omega} (A_{N-8}b^8)^{5/8}=A_{N-8}b^8, 
\end{equation}
and from Bethe equations (\ref{bethe7}),
\begin{equation}
\left[\frac{1}{\omega} (A_{N-8}b^8)^{5/8}\right]^8=\frac{( A_{N-8}b^8)^8}
{b^{5N}}\quad \Rightarrow \quad b^{5N}=1.
\end{equation}

Defining 
\begin{eqnarray}
F_7(w)&=&\prod_{j=1}^{N-8}\prod_{k=0}^{\infty} (1-x^{32k}w/a_j), \\
G_7(1/w)&=&\prod_{j=1}^{N-8}\prod_{k=1}^{\infty} (1-x^{32k}a_j/w),
\end{eqnarray}
and rearranging the Bethe equation (\ref{bethe7.k}) as before, gives
the recurrences
\begin{eqnarray}
\F_7(a)&=&\frac{A(x^{10}a)}{A(x^{22}a)} 
\frac{X(x^{18}a)X^2(x^{20}a)X^2(x^{22}a)X^2(x^{24}a)X(x^{26}a)}
{X(x^6a)X^2(x^{8}a)X^2(x^{10}a)X^2(x^{12}a)X(x^{14}a)} \nonumber \\
&& \times 
\frac{\F_7(x^{10}a)}{\F_7(x^{20}a)},\nonumber \\
&&\nonumber\\
\G_7(1/a)&=&\frac{A(x^{42}/a)}{A(x^{30}/a)} 
\frac{Y(1/x^6a)Y^2(1/x^4a)Y^2(1/x^2a)Y^2(1/a)Y(x^2/a)}
{Y(x^{6}/a)Y^2(x^{8}/a)Y^2(x^{10}/a)Y^2(x^{12}/a)Y(x^{14}/a)}
 \nonumber\\
&& \times\frac{\G_7(x^{10}/a)}{\G_7(x^{20}/a)}.
\end{eqnarray}
The solutions are
\begin{eqnarray}
\F_7(a)&=&\F_0(a) \frac{X(x^{26}a)X(x^{30}a)X(x^{32}a)X(x^{34}a)X(x^{36}a)}
{X(x^{8}a)X(x^{10}a)X(x^{12}a)X(x^{14}a)X(x^{18}a)}
\nonumber \\
&&\times\prod_{m=0}^{\infty}\left\{
\frac{(1-x^{60m+24}\frac{a}{b})(1-x^{60m+28}\frac{a}{b})(1-x^{60m+30}
\frac{a}{b})}
{(1-x^{60m+6}\frac{a}{b})(1-x^{60m+8}\frac{a}{b})(1-x^{60m+10}\frac{a}{b})}
\right.\nonumber\\
&&
\left.\frac{(1-x^{60m+32}\frac{a}{b})(1-x^{60m+34}\frac{a}{b})(1-x^{60m+36}
\frac{a}{b})(1-x^{60m+38}\frac{a}{b})}
{(1-x^{60m+12}\frac{a}{b})(1-x^{60m+16}\frac{a}{b})(1-x^{60m+54}
\frac{a}{b})(1-x^{60m+58}\frac{a}{b})}\right.
\nonumber \\
&&
 \left.\frac{(1-x^{60m+40}\frac{a}{b})(1-x^{60m+42}
\frac{a}{b})(1-x^{60m+46}\frac{a}{b})}{(1-x^{60m+60}\frac{a}{b})
(1-x^{60m+62}\frac{a}{b})(1-x^{60m+64}\frac{a}{b})}\right\}, \label{f7}
\end{eqnarray}
\begin{eqnarray}
\G_7(1/a)&=&\G_0(1/a)
\frac{Y(1/x^4a)Y(1/x^2a)Y(1/a)Y(x^2/a)Y(x^{6}/a)}
{Y(x^{14}/a)Y(x^{18}/a)Y(x^{20}/a)Y(x^{22}/a)Y(x^{24}/a)}
\nonumber \\
&&\times\prod_{m=1}^{\infty}\left\{
\frac{(1-x^{60m-34}\frac{b}{a})(1-x^{60m-32}\frac{b}{a})(1-x^{60m-30}
\frac{b}{a})}
{(1-x^{60m-16}\frac{b}{a})(1-x^{60m-12}\frac{b}{a})(1-x^{60m-10}\frac{b}{a})}
\right.\nonumber\\
&&
\frac{(1-x^{60m-28}\frac{b}{a})(1-x^{60m-24}\frac{b}{a})(1-x^{60m+14}
\frac{b}{a})(1-x^{60m+18}\frac{b}{a})}
{(1-x^{60m-8}\frac{b}{a})(1-x^{60m-6}\frac{b}{a})(1-x^{60m-4}
\frac{b}{a})(1-x^{60m-2}\frac{b}{a})}
\nonumber \\
&&\left. \frac{(1-x^{60m+20}\frac{b}{a})(1-x^{60m+22}
\frac{b}{a})(1-x^{60m+24}\frac{b}{a})}
{(1-x^{60m}\frac{b}{a})(1-x^{60m+2}\frac{b}{a})
(1-x^{60m+6}\frac{b}{a})}\right\}. \label{g7}
\end{eqnarray}
Substitution into 
\begin{eqnarray}
\Lambda_7&=&-w^5 b^3\frac{A(x^{2}w)A(x^{12}w)A(x^{20}/w)A(x^{30}/w)}
{A(x^{2})A(x^{12})A(x^{20})A(x^{30})}\nonumber\\
&&
\times\frac{X(x^{18}w)X(x^{20}w)X(x^{22}w)X(x^{24}w)X(x^{28}w)}
{X(x^{4}w)X(x^{8}w)X(x^{10}w)X(x^{12}w)X(x^{14}w)} \nonumber \\
&&
\times\frac{Y(1/x^{28}w)Y(1/x^{24}w)Y(1/x^{22}w)Y(1/x^{20}w)Y(1/x^{18}w)}
{Y(1/x^{14}w)Y(1/x^{12}w)Y(1/x^{10}w)Y(1/x^{8}w)Y(1/x^{4}w)} \nonumber \\
&& \times\F_7(x^{10}w)\G_7(1/x^{10}w), 
\end{eqnarray}
yields the result (with $b=-1$ and elliptic nome $x^{60}$)
\begin{eqnarray}
\frac{\Lambda_7}{\Lambda_0} &=& w^5
\frac{E(-x^{4}/w)E(-x^{8}/w)E(-x^{10}/w) E(-x^{12}/w)E(-x^{14}/w)}
     {E(-x^{4}w)E(-x^{8}w)E(-x^{10}w) E(-x^{12}w)E(-x^{14}w)}
\nonumber\\
&&\frac{E(-x^{34}w)E(-x^{38}w)
E(-x^{40}w)E(-x^{42}w)E(-x^{44}w)}
{E(-x^{34}/w)E(-x^{38}/w)E(-x^{40}/w)
E(-x^{42}/w)E(-x^{44}/w)} .
\end{eqnarray}
This reduces to 
\begin{equation}
\frac{\Lambda_7}{\Lambda_0} =
\left[
\frac{\tv(\frac{11\pi}{60})\,\tv(\frac{7\pi}{60})
\,\tv(\frac{\pi}{12})\,\tv(\frac{\pi}{20})\,\tv(\frac{\pi}{60})}
     {\tv(\frac{19\pi}{60})\,\tv(\frac{23\pi}{60})
\,\tv(\frac{5\pi}{12})\,\tv(\frac{9\pi}{20})\,\tv(\frac{29\pi}{60})} 
\right]^2, 
\end{equation}
at the isotropic point, with
the elliptic functions of nome $p^{8/15}$.

\subsection{Mass $m_8$}
We begin with the distribution 
$w_j=a_j$ for $j=1, \ldots, N-10$ and $w_{N-9}=b_1x^{-15}$, 
$w_{N-8}=b_2x^{15}$, $w_{N-7}=b_3x^{-13}$,  
$w_{N-6}=b_4x^{13}$, $w_{N-5}=b_5x^{-11}$, $w_{N-4}=b_6x^{11}$,
$w_{N-3}=b_7x^{-9}$, $w_{N-2}=b_8x^{9}$, $w_{N-1}=b_9x^{-7}$,
$w_N=b_{10}x^{7}$. We show in Appendix B.8 that the $b_i$ are equal.
The Bethe equation for the other roots is
\begin{eqnarray}
\lefteqn{-\omega \left[a\frac{E(x^{10}/a)}{E(x^{10}a)}\right]^N=
(A_{N-{10}}b^{10})^{5/8}\frac{a^{10}}{b^{10}}
\frac{E(x^{5}b/a)E^2(x^{7}b/a)E^2(x^{9}b/a)}{E(x^5a/b)E^2(x^7 a/b)
E^2(x^{9} a/b)}} \nonumber \\
&&
\times\frac{E^2(x^{11} b/a)E^2(x^{13}b/a)E(x^{15}b/a)}
{E^2(x^{11}a/b)E^2(x^{13}a/b)E(x^{15}a/b)} 
\prod_{j=1}^{N-10} \frac{E(x^{10}a/a_j)E(x^{20}a_j/a)} 
{E(x^{10}a_j/a)E(x^{20}a/a_j)}. \label{bethe8.k}
\end{eqnarray}
In the $x \to 0$ limit this gives
\begin{equation}
a^{N-10}+\frac{1}{\omega} (A_{N-{10}}b^{10})^{5/8}/b^{10}=0,
\end{equation}
for the $N-10$ zeros. We equate this with
$\prod_{j=1}^{N-10}(a-a_j)$ to obtain
\begin{equation}
 \frac{1}{\omega} (A_{N-{10}}b^{10})^{5/8}=A_{N-10}b^{10}, 
\end{equation}
and from Bethe equations (\ref{bethe8}),
\begin{equation}
\left[\frac{1}{\omega} (A_{N-{10}}b^{10})^{5/8}\right]^{10}
=\frac{(A_{N-{10}}b^{10})^{10}}
{b^{6N}}\quad \Rightarrow \quad b^{6N}=1.
\end{equation}

The functions
\begin{eqnarray}
F_8(w)&=&\prod_{j=1}^{N-10}\prod_{k=0}^{\infty} (1-x^{32k}w/a_j), \\
G_8(1/w)&=&\prod_{j=1}^{N-10}\prod_{k=1}^{\infty} (1-x^{32k}a_j/w),
\end{eqnarray}
must as a consequences of (\ref{bethe8.k}) obey
the recurrences
\begin{eqnarray}
\F_8(a)&=&\frac{A(x^{10}a)}{A(x^{22}a)} 
\frac{X(x^{17}a)X^2(x^{19}a)X^2(x^{21}a)X^2(x^{23}a)X^2(x^{25}a)}
{X(x^5a)X^2(x^{7}a)X^2(x^{9}a)X^2(x^{11}a)X^2(x^{13}a)}
\nonumber \\
&&\times \frac{X(x^{27}a)}{X(x^{15}a)}
         \frac{\F_8(x^{10}a)}{\F_8(x^{20}a)},\\
\G_8(1/a)&=&\frac{A(x^{42}/a)}{A(x^{30}/a)} 
\frac{Y(1/x^7a)Y^2(1/x^5a)Y^2(1/x^3a)}
{Y(x^{5}/a)Y^2(x^{7}/a)Y^2(x^{9}/a)}
 \nonumber\\
&&
\times\frac{Y^2(1/xa)Y^2(x/a)Y(x^3/a)}
{Y^2(x^{11}/a)Y^2(x^{13}/a)Y(x^{15}/a)}
\frac{\G_8(x^{10}/a)}{\G_8(x^{20}/a)}.
\end{eqnarray}
Solving for them, we obtain 
\begin{eqnarray}
\F_8(a)&=&\F_0(a) \frac{X(x^{27}a)X(x^{29}a)X(x^{31}a)X(x^{33}a)X(x^{35}a)
X(x^{37}a)}
{X(x^{7}a)X(x^{9}a)X(x^{11}a)X(x^{13}a)X(x^{15}a)X(x^{17}a)}
\nonumber \\
&&\times\prod_{m=0}^{\infty}\left\{
\frac{(1-x^{60m+25}\frac{a}{b})(1-x^{60m+27}\frac{a}{b})(1-x^{60m+29}
\frac{a}{b})
}
{(1-x^{60m+5}\frac{a}{b})(1-x^{60m+7}\frac{a}{b})(1-x^{60m+9}\frac{a}{b})
}
\right.\nonumber\\
&&\phantom{\times\prod_{m=0}^{\infty}}\left.\frac{(1-x^{60m+31}
\frac{a}{b})(1-x^{60m+33}\frac{a}{b})
(1-x^{60m+35}\frac{a}{b})}{(1-x^{60m+11}\frac{a}{b})
(1-x^{60m+13}\frac{a}{b})(1-x^{60m+15}\frac{a}{b})}\right. \nonumber \\
&&\phantom{\times\prod_{m=0}^{\infty}}\left.
\frac{(1-x^{60m+35}\frac{a}{b})(1-x^{60m+37}\frac{a}{b})(1-x^{60m+39}
\frac{a}{b})}
{(1-x^{60m+55}\frac{a}{b})(1-x^{60m+57}\frac{a}{b})(1-x^{60m+59}\frac{a}{b})}
\right. \nonumber \\
&&\phantom{\times\prod_{m=0}^{\infty}}\left.\frac{(1-x^{60m+41}\frac{a}{b})
(1-x^{60m+43}\frac{a}{b})(1-x^{60m+45}\frac{a}{b})}
{(1-x^{60m+61}\frac{a}{b})(1-x^{60m+63}\frac{a}{b})
(1-x^{60m+65}\frac{a}{b})}\right\}, \label{f8}
\end{eqnarray}
\begin{eqnarray}
\G_8(1/a)&=&\G_0(1/a)
\frac{Y(1/x^5a)Y(1/x^3a)Y(1/xa)Y(x/a)Y(x^3/a)Y(x^{5}/a)}
{Y(x^{15}/a)Y(x^{17}/a)Y(x^{19}/a)Y(x^{21}/a)Y(x^{23}/a)Y(x^{25}/a)}
\nonumber \\
&&\times\prod_{m=1}^{\infty}\left\{
\frac{(1-x^{60m-35}\frac{b}{a})(1-x^{60m-33}\frac{b}{a})
(1-x^{60m-31}\frac{b}{a})}
{(1-x^{60m-15}\frac{b}{a})(1-x^{60m-13}\frac{b}{a})(1-x^{60m-11}\frac{b}{a})}
\right.\nonumber\\
&&\phantom{\times\prod_{m=1}^{\infty}}\left.
\frac{(1-x^{60m-29}\frac{b}{a})(1-x^{60m-27}\frac{b}{a})
(1-x^{60m-25}\frac{b}{a})}
{(1-x^{60m-9}\frac{b}{a})(1-x^{60m-7}\frac{b}{a})(1-x^{60m-5}\frac{b}{a})}
\right.\nonumber \\
&&\phantom{\times\prod_{m=1}^{\infty}}\left.
\frac{(1-x^{60m+15}\frac{b}{a})(1-x^{60m+17}\frac{b}{a})
(1-x^{60m+19}\frac{b}{a})}
{(1-x^{60m-5}\frac{b}{a})(1-x^{60m-3}\frac{b}{a})(1-x^{60m-1}\frac{b}{a})}
\right.\nonumber \\
&&\phantom{\times\prod_{m=1}^{\infty}}\left. \frac{(1-x^{60m+21}\frac{b}{a})(1-x^{60m+23}\frac{b}{a})
(1-x^{60m+25}\frac{b}{a})}
{(1-x^{60m+1}\frac{b}{a})(1-x^{60m+3}\frac{b}{a})
(1-x^{60m+5}\frac{b}{a})}\right\}. \label{g8}
\end{eqnarray}
The eigenvalue expression is
\begin{eqnarray}
\Lambda_8&=&\frac{w^6 }{b^6}\frac{A(x^{2}w)A(x^{12}w)A(x^{20}/w)A(x^{30}/w)}
{A(x^{2})A(x^{12})A(x^{20})A(x^{30})}\nonumber\\
&&
\times\frac{X(x^{17}w)X(x^{19}w)X(x^{21}w)X(x^{23}w)X(x^{25}w)X(x^{27}w)}
{X(x^{5}w)X(x^{7}w)X(x^{9}w)X(x^{11}w)X(x^{13}w)X(x^{15}w)} \nonumber \\
&&
\times\frac{Y(1/x^{27}w)Y(1/x^{25}w)Y(1/x^{23}w)Y(1/x^{21}w)Y(1/x^{19}w)}
{Y(1/x^{15}w)Y(1/x^{13}w)Y(1/x^{11}w)Y(1/x^{9}w)Y(1/x^{7}w)} 
\nonumber \\
&& \times \frac{Y(1/x^{17}w)}{Y(1/x^{5}w)} \, 
\F_8(x^{10}w)\G_8(1/x^{10}w) .
\end{eqnarray}
Thus the leading excitation in the $w^6$ band, with $b=-1$ and
elliptic nome $x^{60}$, is 
\begin{eqnarray}
\frac{\Lambda_8}{\Lambda_0} &=& w^6
\frac{E(-x^{5}/w)E(-x^{7}/w)E(-x^{9}/w)E(-x^{11}/w)}
     {E(-x^{5}w)E(-x^{7}w)E(-x^{9}w)E(-x^{11}w)} \nonumber \\
&&\times
\frac{E(-x^{13}/w)E(-x^{15}/w)E(-x^{35}w)E(-x^{37}w)}
     {E(-x^{13}w)E(-x^{15}w)E(-x^{35}/w)E(-x^{37}/w)} \nonumber\\
&&\times
\frac{E(-x^{39}w)E(-x^{41}w)E(-x^{43}w)E(-x^{45}w)}
     {E(-x^{39}/w)E(-x^{41}/w)E(-x^{43}/w)E(-x^{45}/w)} .\nonumber \\
&&
\end{eqnarray}
At the isotropic point this reduces to
\begin{equation}
\frac{\Lambda_8}{\Lambda_0} =
\left[
\frac{\tv(\frac{\pi}{6})\,\tv(\frac{2\pi}{15})
\,\tv(\frac{\pi}{10})\,\tv(\frac{\pi}{15})\tv(\frac{\pi}{30})\tv(0)}
     {\tv(\frac{\pi}{3})\,\tv(\frac{11\pi}{30})\,
\tv(\frac{2\pi}{5})\,\tv(\frac{13\pi}{30})
\,\tv(\frac{7\pi}{15})\,\tv(\frac{\pi}{2})} 
\right]^2, 
\end{equation}
where the elliptic functions are of nome $p^{8/15}$.

\section{General formula and correlation lengths}

\setcounter{equation}{0}

Having obtained the relevant eigenvalues in the thermodynamic
limit we are now in a position to calculate the correlation
lengths and related mass gaps.

\subsection{Correlation lengths}

Recalling that our transfer matrix acts in the vertical direction,
we consider the pair correlation function between two sites
in the same column of the lattice separated by distance $l$. The
correlation length defining the decay of the correlation function
for large $l$ can be obtained either by
integrating over
the relevant bands of eigenvalues and applying the method of
steepest descent, or equivalently via the leading
eigenvalue at the isotropic point (see, e.g., \cite{Baxter,PB}).
Here we simply follow the latter approach.

Define the quantity
\begin{equation}
r_j(u) = \lim_{N \to \infty} \frac{\Lambda_j(u)}{\Lambda_0(u)}.
\end{equation}
The various correlation lengths follow as
\begin{equation}
\xi_j^{-1} = - \log r_j(u),
\end{equation}
where we are to understand that we take the relevant leading 
eigenvalue at the isotropic point $u = 3\lambda/2$.

\subsection{Regime 1}

For $L$ odd in regime 1, we derived the general result (\ref{genreg1}).
For the leading eigenvalue in the band, our numerical checks confirm that the 
hole is located at $a_N=-1$, with the excitation parameter $b=-1$,
as expected.
Thus using (\ref{gapreg1}) gives the result
\begin{equation}
\xi^{-1} = 2 \log \left[
\frac{\tv(\frac{5\pi}{12},p^{\pi/6\lambda})}
     {\tv(\frac{\pi}{12},p^{\pi/6\lambda})}
\right] . 
\label{correg1}
\end{equation}

The correlation length diverges at criticality, with
\begin{equation}
\xi \sim {1 \over 4 \sqrt 3}\, p^{-\nu_h} \quad \mbox{as} \quad p \to 0,
\end{equation}
where the correlation length exponent $\nu_h$ is given by
\begin{equation}
\nu_h = \frac{r}{6 s} = \frac{2(L+1)}{3 L} .
\label{exp1}
\end{equation}

\subsection{$L=3$ regime 2}

The magnetic Ising value $\lambda=\case{5\pi}{16}$ for $L=3$ in regime 2
has been of particular interest.
Our results for the eight eigenvalue expressions obtained in Section 5
can be summarised in the compact form
\begin{equation}
r_j(w) = w^{n(a)} \prod_a \frac{E(-x^a/w) E(-x^{30+a} w)}
                                {E(-x^a w) E(-x^{30+a}/w)},
\label{compact}
\end{equation}
where the numbers $a$ and $n(a)$ are given in Table 2. 
The $E_8$ numbers $a$ have already appeared in \cite{MO} for the
related Hamiltonian. 
The number $n(a)$ denotes the relevant band of eigenvalues.\\

\begin{table}[ht]
\caption{
Parameters appearing in the eigenvalue expression (\ref{compact}). 
\vskip 5mm
}
\begin{tabular}{||c|c|l||}
\hline
$j$ &  $n(a)$ & $a$ \\
\hline
1   &       2 & 1, 11 \\
2   &       2 & 7, 13 \\
3   &       3 & 2, 10, 12 \\
4   &       3 & 6, 10, 14 \\
5   &       4 & 3, 9, 11, 13 \\
6   &       4 & 6, 8, 10, 12 \\
7   &       5 & 4, 8, 10, 12, 14\\
8   &       6 & 5, 7, 9, 11, 13, 15 \\
\hline
\end{tabular}
\vskip 5mm
\end{table}

In the original notation (\ref{compact}) reads 
\begin{equation}
r_j(u) = \prod_a 
\frac{\tv(\frac{a\pi}{60}-\frac{8 u}{15},p^{8/15}) 
      \td(\frac{a\pi}{60}+\frac{8 u}{15},p^{8/15})} 
     {\td(\frac{a\pi}{60}-\frac{8 u}{15},p^{8/15}) 
      \tv(\frac{a\pi}{60}+\frac{8 u}{15},p^{8/15})},
\label{compactu}
\end{equation}
where $\td(u) = \tv(u+\frac{\pi}{2})$. The zero-momentum excitation energies
of the related Hamiltonian, obtained by inverting the
Fourier transforms in the thermodynamic Bethe Ansatz 
approach \cite{BNW,MO}, are recovered on taking the  
logarithmic derivative of this expression evaluated at $u=0$ and 
applying a Landen transformation.

The first correlation length is given by
\begin{equation}
\xi_1^{-1} = 2 \log \left[
\frac{\tv(\frac{4\pi}{15},p^{8/15})\,\tv(\frac{13\pi}{30},p^{8/15})}
{\tv(\frac{\pi}{15},p^{8/15})\,\tv(\frac{7\pi}{30},p^{8/15})}
\right] . \label{cor1}
\end{equation}
All eight fundamental correlation lengths can be written 
\begin{equation}
m_j = \xi_j^{-1} = 2 \sum_a \log \frac{
\tv(\frac{a\pi}{60}+\frac{\pi}{4},p^{8/15})}
{\tv(\frac{a\pi}{60}-\frac{\pi}{4},p^{8/15})} .
\label{massesp}
\end{equation}
In particular,
\begin{equation}
m_j \sim 8\, p^{8/15}  \sum_a \sin \case{a\pi}{30}
\quad \mbox{as} \quad p \to 0 \, .
\end{equation}
This is the formula obtained by McCoy and Orrick \cite{MO},
from which the $E_8$ masses in (\ref{masses}) are recovered by
virtue of trig identities.

We see that there is surprisingly little variation in the
masses (\ref{massesp}) as a function of the magnetic field-like 
variable $p$. In the limit $p\to1$ the mass ratios are given exactly by
$1 \case{2}{3}$, 2, $2 \case{1}{2}$, 3, $3 \case{1}{3}$, 4, 5, which
are to be compared with the $E_8$ values in (\ref{masses}).

\subsection{Universal magnetic Ising amplitude}

Making use of the Poisson summation formula in the free energy (\ref{fen})
at the magnetic Ising value $\lambda=\case{5\pi}{16}$ we find
\begin{equation}
f \sim 4 \sqrt 3 \, \frac{\sin{\pi \over 5}}{\cos{\pi \over 30}}
                   \, p^{16/15}
\quad \mbox{as} \quad p \to 0 \, .
\end{equation}
On the other hand, from (\ref{cor1}) we have
\begin{equation}
\xi_1 \sim {1 \over 8 \sqrt 3 \, \sin{\pi \over 5}} \, p^{-8/15}
\quad \mbox{as} \quad p \to 0 \, .
\end{equation}
Combining these results gives the universal magnetic Ising amplitude
\begin{equation}
f \, \xi_1^2 = {1 \over 16 \sqrt 3 \sin{\pi \over 5} \cos{\pi \over 30}}
             = 0.061~728~589 \ldots \quad \mbox{as} \quad p \to 0 \, .
\end{equation}
This result has been predicted earlier by other means.
Namely by thermodynamic Bethe Ansatz calculations based on the
$E_8$ scattering theory\cite{Zun,SZ,YZ,F} (see also Ref.~\cite{DM}
in the context of the form-factor bootstrap approach).
Here it is obtained explicitly from the lattice model.

\section{Conclusion}

\setcounter{equation}{0}

We have applied the exact perturbation approach to the
Bethe Ansatz solution of the dilute $A_L$ lattice model 
to derive the free energy per site and the excitation energies 
in regimes 1 and 2 for $L$ odd. 
In the dilute $A_L$ model  
the elliptic nome $p$ is magnetic field-like for $L$ odd 
and is temperature-like for $L$ even. The particular point
in regime 2 for $L=3$ has attracted considerable recent
attention, being in the same universality class as the
Ising model in a magnetic field. We have specifically considered
the case $L$ odd for which the method perturbs from the
ordered high field limit. Our result for the free energy 
(\ref{fen}) is in agreement with that obtained via the
inversion relation method for general $L$ \cite{WNS,WPSN}.

In regime 1 the leading excitations in the $w$ band are
1-strings, characterised by the string excitation parameter $b$
and a single hole. Our final result for the related correlation
length is given in (\ref{correg1}). The excitations are
considerably more complicated in regime 2, where we have
concentrated on the case $L=3$. Here our results for the leading 
eigenvalue associated with each mass are 
summarised in (\ref{compact}) and (\ref{compactu}).  
The corresponding inverse correlation lengths or masses
are given in (\ref{massesp}), all of which give the
magnetic Ising correlation length exponent $\nu_h = \frac{8}{15}$.
In particular, the $E_8$ masses (\ref{masses}) predicted by
Zamolodchikov appear with the approach to criticality.

Of course these masses have been obtained earlier in the
scaling limit via the thermodynamic Bethe Ansatz approach \cite{BNW}. 
Although the calculations are somewhat complicated our approach
nevertheless provides in principle a means of classifying
and counting all excitations in the eigenvalue bands.
We have not pursued this classification in detail, rather
contenting ourselves with the location of the leading eigenvalue
in each band, of relevance to the correlation lengths.
As for the thermodynamic Bethe Ansatz, our calculations
rely upon the key input of the string configurations.
We have taken as our starting point the eight thermodynamically
significant string excitation types revealed in the previous numerical 
studies \cite{BNW,GNa,GN}. The one exception is that we have
considered the 7-string configuration \cite{GN} for mass $m_4$ 
(see Table 1) rather than the 5-string \cite{BNW}. However, although
we have not included here our calculations based on the 5-string,
we find that the 5-string configuration is also consistent in the
ordered limit. Moreover, we find that both the 5-string and the 
7-string lead to
the \emph{same} eigenvalue expression and thus mass. The
origin of this behaviour is seen in the 7-string calculation
where the string parameters $\alpha$ and $\beta$ (see eqn (B.30)
do not appear in the final result (\ref{reg24b}). In both cases, 
it is the unpaired string at $b x^{16}$ which ultimately drives
the calculation.

The derivation of the correlation length for $L\ne3$ in regime 2
is complicated. In this regime the leading excitation in the $w$
band has $\ell=\case{1}{2}(L-1)+1$ and, like the leading 2-string
in the $w^2$ band for $L=3$, it begins life for small $N$ and
$p \simeq0$ as a 1-string. We have not pursued this further.
Nevertheless we have numerically
observed that the final result (\ref{reg1}) for regime 1 also 
applies to the leading $w$ band excitation in regime 2. We thus 
believe that the correlation length (\ref{cor1}) and the corresponding
exponents 
\begin{equation}
\nu_h = \frac{r}{6 s} =
\frac{2(L+1)}{3(L+2)}  
\label{exp2}
\end{equation}
also hold in regime 2 for $L \ne 3$. This result includes 
$\nu_h = \case{8}{15}$ for $L=3$.

The correlation length exponents are seen to satisfy the
general scaling relation
$2 \nu_h = 1 + 1/\delta$, which follows from the general relation
$f \, \xi^2 \sim \mbox{constant}$,
where $f \sim p^{1+ 1/\delta}$ is the singular part of the bulk
free energy and the exponents $\delta$ are those following from
the singular behaviour of (\ref{fen})
\cite{WNS,WPSN}. The same correlation length exponents
should hold for $L$ even, for which the integrable perturbation is
thermal-like. The scaling relation
is now $2\nu_t = 2 - \alpha$, where $\nu_t$ is as given in
(\ref{exp1}), (\ref{exp2}) and $\alpha$ in \cite{WNS,WPSN}. 
In particular,
(\ref{exp1}) gives the Ising value $\nu_t=1$ for $L=2$ in regime 1,
as expected.

Another approach to the results given here 
is to use the inversion relation method,
as was originally used to obtain the bulk \cite{WNS,WPSN}
and excess surface \cite{BFZ} free energies. This approach
can also be applied to the excitations.\footnote{Yet another
would involve integral equations for the root densities,
as done for the eight-vertex model \cite{JKM}.} 
This has been done, for
example, for the eight-vertex \cite{KZ} and the 
Andrews-Baxter-Forrester models \cite{OP}.
Although in principle completely avoiding the string hypothesis
the inversion relation method
still requires strong assumptions on the analyticity properties
of the eigenvalues. Our explicit results for $r_j(u)$ give these analyticity
properties {\em a posteriori}. It remains to be seen if
these properties can be established {\em a priori}, thus allowing an
alternative derivation of the mass gaps.
We note that the inversion relation
\begin{equation}
r_j(u) \, r_j(u+3\lambda) = 1
\end{equation}
is indeed satisfied by our results. Alternatively, 
$r_j(w) \, r_j(x^{30} w) = 1$ for $L=3$ in regime 2, which
is seen to hold trivially in view of (\ref{compact}).
There is a further relation\footnote{We thank 
A. Kl\"umper for pointing this relation out to us.} 
\begin{equation}
r_j(u) \, r_j(u+2\lambda)=r_j(u+\lambda),
\end{equation}
which is also easily seen to be satisfied by our results.

\ack
It is a pleasure to thank V. Bazhanov, V. Dunjko, U. Grimm, 
A. Kl\"umper, B.M. McCoy, S.O. Warnaar and A.B. Zamolodchikov 
for some helpful remarks. This work was undertaken while KAS was a 
visitor at the Center for Mathematics and Its Applications at The
Australian National University, which was facilitated by a Commonwealth
Staff Development Fund grant administered by the Academic Development
Unit of La Trobe University.
The work of MTB has been supported by the Australian Research Council.
\newpage
\renewcommand{\thesection}{\Alph{section}}
\renewcommand{\theequation}{\Alph{section}.\arabic{equation}}
\section*{Appendices}
\setcounter{section}{0}
\section{Auxiliary functions}

\setcounter{equation}{0}

The simplest identities for the auxiliary functions (4.6)-(4.8) are
\begin{eqnarray}
\frac{X(x^nw)}{X(x^{n+2r}w)}&=&(1-x^n w/b),\\
\frac{Y(x^{n-2r}/w)}{Y(x^n/w)}&=&(1-x^nb/w),\\
\frac{R(x^nw)}{R(x^{n+2r}w)}&=&(1-x^n w/a_N),\\
\frac{S(x^{n-2r}/w)}{S(x^n/w)}&=&(1-x^n a_N/w).
\end{eqnarray}

When solving the recurrence relations (in either regime), their
generalisations, 
\begin{eqnarray}
\prod_{m=0}^{\infty}\frac{X(x^{12ms+n}w)}
{X(x^{12ms+n+2r}w)}&=&
\prod_{m=0}^{\infty}(1-x^{12ms+n}w/b),\\
\prod_{m=0}^{\infty}\frac{Y(x^{12ms+n-2r}/w)}
{Y(x^{12ms+n}/w)}&=&
\prod_{m=0}^{\infty}(1-x^{12ms+n}b/w),
\end{eqnarray}
are used. In regime 1 we also require
\begin{eqnarray}
\prod_{m=0}^{\infty}\frac{R(x^{12ms+n}w)}
{R(x^{12ms+n+2r}w)}&=&
\prod_{m=0}^{\infty}(1-x^{12ms+n}w/a_N),\\
\prod_{m=0}^{\infty}\frac{S(x^{12ms+n-2r}/w)}
{S(x^{12ms+n}/w)}&=&
\prod_{l=0}^{\infty}(1-x^{12ms+n}a_N/w).
\end{eqnarray}
In regime 2 we treat only $L=3$, so that $s=5$ and
$r=16$.  For mass $m_4$ we need the analogous identities,
\begin{eqnarray}
\prod_{m=0}^{\infty}\frac{R_4(x^{60m+n}w)}
{R_4(x^{60m+n+32}w)}&=&
\prod_{m=0}^{\infty}(1-x^{60m+n}w/a_{N-6}),\\
\prod_{m=0}^{\infty}\frac{S_4(x^{60m+n-32}/w)}
{S_4(x^{60m+n}/w)}&=&
\prod_{l=0}^{\infty}(1-x^{12ms+n}a_{N-6}/w).
\end{eqnarray}
Of course, the regime 2 auxiliary functions $X_{\alpha}(w)$ and 
$X_{\beta}(w)$ obey the same identities as $X(w)$, with $b$ therein 
replaced by $\alpha$ and $\beta$, respectively. Similarly
$Y_{\alpha}(1/w)$ and $Y_{\beta}(1/w)$ obey the same identities as 
$Y(1/w)$, with again $b$ replaced by
$\alpha$ and $\beta$ as appropriate.

When substituting $\F$ and $\G$ into the $k=N$ Bethe equation in 
regime 1 we use
\begin{eqnarray}
\prod_{l=0}^{\infty}\frac{X(x^{(2r-4s)l+n}w)}
{X(x^{(2r-4s)l+n+2r}w)}&=&
\prod_{l=0}^{\infty}(1-x^{(2r-4s)l+n}w/b),\\
\prod_{l=0}^{\infty}\frac
{Y(x^{(2r-4s)l+n-2r}/w)}{Y(x^{(2r-4s)l+n}/w)}&=&
\prod_{l=0}^{\infty}(1-x^{(2r-4s)l+n}b/w),\\
\prod_{l=0}^{\infty}\frac{R(x^{(2r-4s)l+n}w)}
{R(x^{(2r-4s)l+n+2r}w)}&=&
\prod_{l=0}^{\infty}(1-x^{(2r-4s)l+n}w/a_N),\\
\prod_{l=0}^{\infty}\frac{S(x^{(2r-4s)l+n-2r}/w)}
{S(x^{(2r-4s)l+n}/w)}&=&
\prod_{l=0}^{\infty}(1-x^{(2r-4s)l+n}a_N/w).
\end{eqnarray}

To combine the Bethe equations from the strings in Appendix B, once
the functions $\F_i$ and $\G_i$ have been found, we use
\begin{eqnarray}
\prod_{m=0}^{\infty}\frac{A(x^{60m+n}/b)}
{A(x^{60m+n+32}/b)}&=&
\prod_{m=0}^{\infty}(1-x^{60m+n}/b)^N,\\
\prod_{m=0}^{\infty}\frac{A(x^{60m+n}b)}
{A(x^{60m+n+32}b)}&=&
\prod_{m=0}^{\infty}(1-x^{60m+n}b)^N,\\
\prod_{m=0}^{\infty}\frac{A(x^{60m+60+n}/b)}
{A(x^{60m+n+32}/b)}&=&
X(x^n)^{-N}\prod_{m=0}^{\infty}(1-x^{60m+60+n}/b)^N,\\
\prod_{m=0}^{\infty}\frac{A(x^{60m+60+n}b)}
{A(x^{60m+n+32}b)}&=&
Y(x^n)^{-N}\prod_{m=0}^{\infty}(1-x^{60m+60+n}b)^N.
\end{eqnarray}
These last four relations are just special cases of more general 
identities (which we do not require)
which would have $12s$ in place of $60$ and $2r$ in place of $32$.

\section{Bethe equations for the masses}

\setcounter{equation}{0}

In this Appendix we consider the Bethe equations for the 
eight string types corresponding to the $E_8$ masses. Their
chief usefulness is to establish analytically that the
coefficient of each member of a string is the same. (In the
case of $m_4$ and $m_6$, these equations give equality only
between subsets of the coefficients, and this feature
was explicitly included in the calculations in the body of
the paper.) 
Secondly, once the auxiliary functions (which give the remainder
of the Bethe roots, the $a_k$) are substituted into the 
string equations, it is possible to establish a condition on
the string coefficient $b$ in which the same patterns in the exponents
of $x$ arise as do in the eigenvalues (or masses) themselves. 
Such higher level Bethe equations for the string parameters have been 
discussed, for example, for the CSOS model \cite{PB}. In each
case these equations ensure that the corresponding eigenvalue expression 
reduces to an $N$th root of unity at $w=1$. This must be so, as the
row transfer matrix reduces to a shift operator at this point. 

Because some of the strings are quite long, it will
be convenient to introduce the notation
$\prod_{i=1}^{m}b_i=B_m$.

\subsection{Mass $m_1$} \label{b1}

With the roots $w_j=a_j$ for $j=1, \ldots, N-2$ and 
$w_{N-1}=b_1x^{-11}$, $w_N=b_2x^{11}$ the Bethe equations for 
$k=N-1$ and $k=N$ are
\begin{eqnarray}
\lefteqn{-\omega \left[x^{-10}\frac{E(x^{21}/b_1)}{E(x/b_1)}\right]^N=
(A_{N-2}B_2)^{-3/8}b_1^{N}x^{-N-18}}
\nonumber
\\ &&\times
\frac{E(x^{10} b_2/b_1)E(x^{12}b_2/b_1)}{E(b_1/b_2)E(x^{2} b_2/b_1)} 
\prod_{j=1}^{N-2} \frac{E(xa_j/b_1)E(x^{31}a_j/b_1)} 
{E(x^{21}a_j/b_1)E(x^{23}a_j/b_1)}, \label{bethe1.1}\\
&& \nonumber\\
&&\nonumber\\
\lefteqn{-\omega \left[x^{10}\frac{E(xb_2)}{E(x^{21}b_2)}\right]^N=
(A_{N-2}B_2)^{-3/8}b_2^{N}x^{N+18}}
\nonumber
\\ &&\times
\frac{E(b_1/b_2)E(x^{2} b_2/b_1)}{E(x^{10} b_2/b_1)E(x^{12}b_2/b_1)}
\prod_{j=1}^{N-2} \frac{E(x^{21}b_2/a_j)E(x^{9}a_j/b_2)} 
{E(xa_j/b_2)E(x^{31}a_j/b_2)}. \label{bethe1.2}
\end{eqnarray}
Taken individually as $x \rightarrow 0$, these equations imply that 
\begin{equation}
(1-\frac{b_1}{b_2})=O(x^{9N})
\Rightarrow\frac{b_1}{b_2}=1+O(x^{9N}).
\end{equation}
Thus in the thermodynamic ($N \rightarrow \infty$) limit
$b_1=b_2=b$.

Forming the product of (\ref{bethe1.1}) and (\ref{bethe1.2}), we see 
that many factors cancel and those remaining can be written in terms of
the auxiliary functions (\ref{def1a}) and (\ref{def1b}) for $m_1$, 
namely 
\begin{equation}
\left[\frac{E(xb)E(x^{21}/b)}{E(x/b)E(x^{21}b)}\right]^N=
\frac{b^{2N}}
{\F_1(x^9b)\F_1(x^{11}b)\G_1(1/x^9b)\G_1(1/x^{11}b)}.
\end{equation}
Using the functions (\ref{f1}) and (\ref{g1}) and 
identities from Appendix A, we obtain
\begin{equation}
\left[\frac{E(xb,x^{60})E(x^{11}b,x^{60})E(x^{31}/b,x^{60})E(x^{41}/b,x^{60})}
{E(x/b,x^{60})E(x^{11}/b,x^{60})E(x^{31}b,x^{60})E(x^{41}b,x^{60})}\right]^N
=b^{2N}.
\end{equation}
This defining relationship is the higher level Bethe equation to be
satisfied by the string parameter $b$. Note that it is trivially 
satisfied by $b=-1$.
Compare also the pattern of exponents in this equation
with the expression (\ref{reg21b}) for $\Lambda_1$.
Taken together they ensure that the eigenvalue is an
$N$th root of unity at $w=1$.

\subsection{Mass $m_2$} \label{b2}
With the roots
$w_j=a_j$ for $j=1, \ldots, N-4$ and $w_{N-3}=b_1x^{-5}$, 
$w_{N-2}=b_2x^{5}$, $w_{N-1}=b_3x^{-15}$, $w_N=b_4x^{15}$ the Bethe 
equations for $k=N-3, N-2, N-1,N$, respectively,  are
\begin{eqnarray}
\lefteqn{-\omega \left[b_1x^{-5}\frac{E(x^{15}/b_1)}{E(x^5 b_1)}\right]^N=
(A_{N-4}B_4)^{5/8}b_1^{2}b_4^{-2}x^{-18}}
\nonumber
\\ &&\times
\frac{E(b_1/b_2)E(x^{30}b_2/b_1)E(x^{10}b_3/b_1)E(x^{20}b_1/b_3)
E(x^{8}b_4/b_1)E(x^{10}b_4/b_1)}
{E(x^{10}b_1/b_2)E(x^{20}b_2/b_1)E(b_3/b_1)E(x^{30}b_1/b_3)
E(b_1/b_4)E(x^{30}b_4/b_1)}
\nonumber
\\ &&\times
 \prod_{j=1}^{N-4} \frac{E(x^5b_1/a_j)E(x^{25}a_j/b_1)} 
{E(x^{15}a_j/b_1)E(x^{15}b_1/a_j)}, \label{bethe2.1}\\
&&\nonumber\\
&&\nonumber\\
\lefteqn{-\omega \left[b_2x^{5}\frac{E(x^{5}/b_2)}{E(x^{15} b_2)}\right]^N=
(A_{N-4}B_4)^{5/8}b_2^{2}b_3^{-2}x^{18}}
\nonumber
\\ &&\times
\frac{E(x^{10}b_1/b_2)E(x^{20}b_2/b_1)E(b_3/b_2)E(x^{30}b_2/b_3)
E(b_2/b_4)E(x^{30}b_4/b_2)}
{E(b_1/b_2)E(x^{30}b_2/b_1)E(x^{8}b_2/b_3)E(x^{10}b_2/b_3)
E(x^{10}b_2/b_4)E(x^{20}b_4/b_2)}
\nonumber
\\ &&\times
 \prod_{j=1}^{N-4} \frac{E(x^{15}b_2/a_j)E(x^{15}a_j/b_2)} 
{E(x^{5}a_j/b_2)E(x^{25}b_2/a_j)}, \label{bethe2.2}\\
&&\nonumber\\
&&\nonumber\\
\lefteqn{-\omega \left[x^{-10}\frac{E(x^{25}/b_3)}{E(x^{5}/ b_3)}\right]^N=
(A_{N-4}B_4)^{5/8}A_{N-4}^{-2}b_3^{2N-6}b_2^{-2}x^{-8N-6}}
\nonumber
\\ &&\times
\frac{E(b_3/b_1)E(x^{30}b_1/b_3)E(x^{8}b_2/b_3)E(x^{10}b_2/b_3)
E(x^{20}b_4/b_3)E(x^{18}b_4/b_3)}
{E(x^{10}b_3/b_1)E(x^{20}b_1/b_3)E(b_3/b_2)E(x^{30}b_2/b_3)
E(x^{8}b_4/b_3)E(x^{10}b_4/b_3)}
\nonumber
\\ &&\times
 \prod_{j=1}^{N-4} \frac{E(x^{3}a_j/b_3)E(x^{5}a_j/b_3)} 
{E(x^{25}a_j/b_3)E(x^{5}b_3/a_j)}, \label{bethe2.3}\\
&&\nonumber\\
&&\nonumber\\
\lefteqn{-\omega \left[x^{10}\frac{E(x^{5}b_4)}{E(x^{25}b_4)}\right]^N=
(A_{N-4}B_4)^{5/8}A_{N-4}^{-2}b_4^{2N-6}b_1^{-2}x^{8N+6}}
\nonumber
\\ &&\times
\frac{E(b_1/b_4)E(x^{30}b_4/b_1)E(x^{10}b_2/b_4)E(x^{20}b_4/b_2)
E(x^{8}b_4/b_3)E(x^{10}b_4/b_3)}
{E(x^{8}b_4/b_1)E(x^{10}b_4/b_1)E(b_2/b_4)E(x^{30}b_4/b_2)
E(x^{20}b_4/b_3)E(x^{18}b_4/b_3)}
\nonumber
\\ &&\times
 \prod_{j=1}^{N-4} \frac{E(x^{3}a_j/b_4)E(x^{5}a_j/b_4)} 
{E(x^{25}a_j/b_4)E(x^{5}b_4/a_j)}. \label{bethe2.4}
\end{eqnarray}
Taken individually as $x \rightarrow 0$, these equations imply that
\begin{eqnarray}
\frac{(1-\frac{b_3}{b_1})(1-\frac{b_1}{b_4})}
{(1-\frac{b_1}{b_2})}&=&O(x^{5N}), \qquad \frac{(1-\frac{b_3}{b_2})}
{(1-\frac{b_3}{b_1})}=O(x^{2N}), \nonumber \\
\frac{(1-\frac{b_3}{b_2})(1-\frac{b_2}{b_4})}
{(1-\frac{b_1}{b_2})}&=&O(x^{5N}), \qquad
\frac{(1-\frac{b_1}{b_4})}
{(1-\frac{b_2}{b_4})}=O(x^{2N}). 
\end{eqnarray}
We cannot read off the relationship between the $b_i$ in this case, as we
could for $m_1$, but we see that these relations are satisfied if
\begin{equation}
b_1=b_4+O(x^{5N}),\quad
b_1=b_2+O(x^{3N}),\quad
b_1=b_3+O(x^{3N}).
\end{equation}
Thus we conclude $b_1=b_2=b_3=b_4=b$ and $B_4=b^4$.

The product of the four Bethe equations can be written in terms of
the auxiliary functions (\ref{def2}) for $m_2$, with 
\begin{eqnarray}
\lefteqn{\left[\frac{E(x^{15}/b)E(x^{25}/b)}{E(x^{15}b)E(x^{25}b)}
\right]^N=} \nonumber \\
&&
b^{2N}\left[\F_2(x^3b)\F_2(x^5b)\F_2(x^{15}b)\F_2(x^{17}b)
\right]^{-1}\nonumber
\\ &&\times\left[\G_2(1/x^3b)\G_2(1/x^5b)\G_2(1/x^{15}b)
\G_2(1/x^{17}b)\right]^{-1}. 
\end{eqnarray}
Using the functions (\ref{f2}) and (\ref{g2}) and identities from 
Appendix A, we obtain
\begin{equation}
\left[\frac{E(x^7b,x^{60})E(x^{13}b,x^{60})E(x^{37}/b,x^{60})E(x^{43}/b,x^{60})}
{E(x^7/b,x^{60})E(x^{13}/b,x^{60})E(x^{37}b,x^{60})E(x^{43}b,x^{60})}\right]^N
=b^{2N}.
\end{equation}

\subsection{Mass $m_3$} \label{b3}
With the roots
$w_j=a_j$ for $j=1, \ldots, N-4$ and $w_{N-3}=b_1x^{-10}$, 
$w_{N-2}=b_2x^{10}$, $w_{N-1}=b_3x^{-20}$, $w_N=b_4x^{20}$ the Bethe 
equations for $k=N-3, N-2, N-1,N$ read 
\begin{eqnarray}
\lefteqn{-\omega \left[b_1x^{-10}\frac{E(x^{20}/b_1)}{E( b_1)}\right]^N=
(A_{N-4}B_4)^{5/8}b_1^{2}b_2^{-2}x^{-38}}
\nonumber
\\ &&\times
\frac{E(x^8 b_2/b_1)E(x^{10}b_2/b_1)E(x^{10}b_3/b_1)E(x^{20}b_1/b_3)
E(x^{18}b_4/b_1)E(x^{20}b_4/b_1)}
{E(b_1/b_2)E(x^{30}b_2/b_1)E(b_3/b_1)E(x^{30}b_1/b_3)
E(x^8b_4/b_1)E(x^{10}b_4/b_1)} 
\nonumber
\\ &&\times
 \prod_{j=1}^{N-4} \frac{E(b_1/a_j)E(x^{30}a_j/b_1)} 
{E(x^{20}a_j/b_1)E(x^{10}b_1/a_j)}, \label{bethe3.1}\\
&&\nonumber\\
&&\nonumber\\
\lefteqn{-\omega \left[x^{10}\frac{E(b_2)}{E(x^{20} b_2)}\right]^N=
(A_{N-4}B_4)^{5/8}b_2^{2}b_1^{-2}x^{38}}
\nonumber
\\ &&\times
\frac{E(b_1/b_2)E(x^{30}b_2/b_1)E(x^8 b_2/b_3)E(x^{10}b_2/b_3)
E(b_2/b_4)E(x^{30}b_4/b_2)}
{E(x^8 b_2/b_1)E(x^{10}b_2/b_1)E(x^{18}b_2/b_3)E(x^{20}b_2/b_3) 
E(x^{10}b_2/b_4)E(x^{20}b_4/b_2)} 
\nonumber
\\ &&\times
 \prod_{j=1}^{N-4} \frac{E(x^{20}b_2/a_j)E(x^{10}a_j/b_2)} 
{E(a_j/b_2)E(x^{30}b_2/a_j)}, \label{bethe3.2}\\
&&\nonumber\\
&&\nonumber\\
\lefteqn{-\omega \left[x^{-10}\frac{E(x^{30}/b_3)}{E(x^{10}/ b_3)}\right]^N=
(A_{N-4}B_4)^{5/8}A_{N-4}^{-2}b_3^{2N-8}x^{-18N+32}}
\nonumber
\\ &&\times
\frac{E(b_3/b_1)E(x^{30}b_1/b_3)E(x^{18}b_2/b_3)E(x^{20}b_2/b_3)
E(x^{28}b_4/b_3)E(x^{30}b_4/b_3)}
{E(x^{10}b_3/b_1)E(x^{20}b_1/b_3)E(x^8 b_2/b_3)E(x^{10}b_2/b_3)
E(x^{18}b_4/b_3)E(x^{20}b_4/b_3)} 
\nonumber
\\ &&\times
 \prod_{j=1}^{N-4} \frac{E(x^{10}a_j/b_3)E(x^{8}a_j/b_3)} 
{E(x^{30}a_j/b_3)E(b_3/a_j)}, \label{bethe3.3}\\
&&\nonumber\\
&&\nonumber\\
\lefteqn{-\omega \left[x^{10}\frac{E(x^{10}b_4)}{E(x^{30}b_4)}\right]^N=
(A_{N-4}B_4)^{5/8}A_{N-4}^{-2}b_4^{2N-8}x^{18N-32}}
\nonumber
\\ &&\times
\frac{E(x^8 b_4/b_1)E(x^{10}b_4/b_1)E(x^{10}b_2/b_4)E(x^{20}b_4/b_2)
E(x^{18}b_4/b_3)E(x^{20}b_4/b_3)}
{E(x^{18}b_4/b_1)E(x^{20}b_4/b_1)E(b_2/b_4)E(x^{30}b_4/b_2)
E(x^{28}b_4/b_3)E(x^{30}b_4/b_3)} 
\nonumber
\\ &&\times
 \prod_{j=1}^{N-4} \frac{E(x^{30}b_4/ba_j)E(a_j/b_4)} 
{E(x^{8}b_4/a_j)E(x^{10}b_4/a_j)}. \label{bethe3.4}
\end{eqnarray}
Taken individually as $x \rightarrow 0$, these equations imply that
\begin{eqnarray}
(1-\frac{b_1}{b_2})(1-\frac{b_3}{b_1})
&=&O(x^{10N}), \qquad
(1-\frac{b_3}{b_1})
=O(x^{8N}), \nonumber \\
(1-\frac{b_1}{b_2})(1-\frac{b_2}{b_4})&=&O(x^{10N}), \qquad
(1-\frac{b_2}{b_4})=O(x^{8N}), 
\end{eqnarray}
from which we can read off
\begin{equation}
b_1=b_3+O(x^{8N}), \quad b_2=b_4+O(x^{8N}),
\quad b_1=b_2+O(x^{2N}).
\end{equation}
Thus we conclude $b_1=b_2=b_3=b_4=b$ and $B_4=b^4$.

Expressing the product of the four Bethe equations in terms of
the $m_3$ auxiliary functions (\ref{def3}) gives
\begin{eqnarray}
\lefteqn{\left[\frac{E(x^{20}/b)E(x^{22}/b)E(x^{30}/b)}
{E(x^{20}b)E(x^{22}b)E(x^{30}b)}\right]^N=}
\nonumber \\
&&b^{3N}
\left[\F_3(x^8b)\F_3^2(x^{10}b)\F_3(x^{12}b)
\G_3(1/x^8 b)\G_3^2(1/x^{10}b)\G_3(1/x^{12}b)\right]^{-1}.
\end{eqnarray}
Using the functions (\ref{f3}) and (\ref{g3}) and 
identities from Appendix A we obtain, in terms of elliptic functions 
(of nome $x^{60}$, \emph{not} $x^{32}$),
\begin{equation}
\left[\frac{E(x^2b)E(x^{10}b)E(x^{12}b)
E(x^{32}/b)E(x^{40}/b)E(x^{42}/b)}
{E(x^2/b)E(x^{10}/b)E(x^{12}/b)
E(x^{32}b)E(x^{40}b)E(x^{42}b)}\right]^N=b^{3N}.
\end{equation}

\subsection{Mass $m_4$} \label{b4}

With the roots
$w_j=a_j$ for $j=1, \ldots, N-7$ and 
$w_{N-6}=b_1x^{-1}$, $w_{N-5}=b_2x$, $w_{N-4}=b_3x^{-7}$,
$w_{N-3}=b_4x^{7}$, $w_{N-2}=b_5x^{-13}$, $w_{N-1}=b_6x^{13}$, 
$w_N=b_7x^{16}$ the Bethe equations for $k=N-6,\ldots,
N$ are
\begin{eqnarray}
\lefteqn{-\omega \left[b_1x^{-1}\frac{E(x^{11}/b_1)}{E(x^9 b_1)}\right]^N=
(A_{N-7}B_7)^{5/8}b_1^{6}(b_5b_6b_7)^{-2}x^{-6}}
\nonumber
\\ &&\times
\frac{E(x^8 b_1/b_2)E(x^{10}b_1/b_2)E(x^{14}b_3/b_1)E(x^{16}b_1/b_3)
E(x^{2}b_4/b_1)E(x^{28}b_4/b_1)}
{E(x^{12}b_2/b_1)E(x^{18}b_1/b_2)E(x^4 b_3/b_1)E(x^{26}b_1/b_3)
E(x^{18} b_4/b_1)E(x^{20}b_4/b_1)}
\nonumber
\\ &&\times
\frac{E(x^{8}b_5/b_1)E(x^{10}b_5/b_1)E(x^{4}b_6/b_1)E(x^{2}b_6/b_1)
E(x^{7}b_7/b_1)E(x^{5}b_7/b_1)}
{E(b_1/b_5)E(x^{2} b_1/b_5)E(x^{6} b_1/b_6)E(x^{8}b_1/b_6)
E(x^{3} b_1/b_7)E(x^{5}b_1/b_7)}
\nonumber
\\ &&\times
 \prod_{j=1}^{N-7} \frac{E(x^9 b_1/a_j)E(x^{21}a_j/b_1)} 
{E(x^{11}a_j/b_1)E(x^{19}b_1/a_j)}, \label{bethe4.1}\\
&&\nonumber\\
&&\nonumber\\
\lefteqn{-\omega \left[b_2x\frac{E(x^{9}/b_2)}{E(x^{11} b_2)}\right]^N=
(A_{N-7}B_7)^{5/8}b_2^{6}(b_5b_6b_7)^{-2}x^{6}}
\nonumber\\ 
&&\times
\frac{E(x^{12} b_2/b_1)E(x^{18}b_1/b_2)E(x^{18}b_2/b_3)E(x^{12}b_3/b_2)
E(x^{4}b_2/b_4)E(x^{26}b_4/b_2)}
{E(x^{8}b_1/b_2)E(x^{10}b_1/b_2)E(x^2 b_3/b_2)E(x^{28}b_2/b_3)
E(x^{16} b_4/b_2)E(x^{18}b_4/b_2)}
\nonumber \\ &&\times
\frac{E(x^{8}b_2/b_5)E(x^{6}b_5/b_2)E(x^{2}b_6/b_2)E(b_2/b_6)
E(x^{5}b_7/b_2)E(x^{3}b_7/b_2)}
{E(x^4 b_2/b_5)E(x^{2} b_2/b_5)E(x^{8}
b_2/b_6)E(x^{10}b_2/b_6) E(x^{7} b_2/b_7)E(x^{5}b_2/b_7)}
\nonumber \\ 
&&\times
\prod_{j=1}^{N-7} \frac{E(x^{11}b_2/a_j)E(x^{19}a_j/b_2)} 
{E(x^{9}a_j/b_2)E(x^{21}b_2/a_j)}, \label{bethe4.2}\\
&&\nonumber\\
&&\nonumber\\
\lefteqn{-\omega \left[b_3x^{-7}\frac{E(x^{17}/b_3)}{E(x^{3} b_3)}\right]^N=
(A_{N-7}B_7)^{5/8}b_3^{4}(b_4b_6)^{-2}x^{-34}}
\nonumber\\ 
&&\times
\frac{E(x^4 b_3/b_1)E(x^{26}b_1/b_3)E(x^2 b_3/b_2)E(x^{28}b_2/b_3)
E(x^{4}b_4/b_3)E(x^{2}b_4/b_3)}
{E(x^{14}b_3/b_1)E(x^{16}b_1/b_3)E(x^{18}b_2/b_3)E(x^{12}b_3/b_2)
E(x^{6} b_3/b_4)E(x^{8}b_3/b_4)} 
\nonumber \\ &&\times
\frac{E(x^{14}b_5/b_3)E(x^{16}b_5/b_3)E(x^{10}b_6/b_3)E(x^8 b_6/b_3)
E(x^{13}b_7/b_3)E(x^{11}b_7/b_3)}
{E(x^4 b_3/b_5)E(x^{6} b_5/b_3)E(x^{2} b_3/b_6)E(b_3/b_6)
E(xb_7/b_3)E(x^{3}b_7/b_3)}
\nonumber \\ 
&&\times
\prod_{j=1}^{N-7} \frac{E(x^{3}b_3/a_j)E(x^{27}a_j/b_3)} 
{E(x^{17}a_j/b_3)E(x^{13}b_3/a_j)}, \label{bethe4.3}\\
&&\nonumber\\
&&\nonumber\\
\lefteqn{-\omega \left[b_4x^{7}\frac{E(x^{3}/b_4)}{E(x^{17} b_4)}\right]^N=
(A_{N-7}B_7)^{5/8}b_4^{4}(b_5b_3)^{-2}x^{34}}
\nonumber\\ 
&&\times
\frac{E(x^{18} b_4/b_1)E(x^{20}b_4/b_1)E(x^{16} b_4/b_2)E(x^{18}b_4/b_2)
E(x^{6} b_3/b_4)E(x^{8}b_3/b_4)}
{E(x^{2}b_4/b_1)E(x^{28}b_4/b_1)E(x^{4}b_2/b_4)E(x^{26}b_4/b_2)
E(x^{4}b_4/b_3)E(x^{2}b_4/b_3)}
\nonumber \\ &&\times
\frac{E(x^{2}b_5/b_4)E(b_5/b_4)E(x^{4}b_4/b_6)E(x^6 b_4/b_6)
E(xb_4/b_7)E(x^{3}b_4/b_7)}
{E(x^{10} b_4/b_5)E(x^{8} b_4/b_5)E(x^{16} b_4/b_6)E(x^{14}b_4/b_6)
E(x^{11} b_4/b_7)E(x^{13}b_4/b_7)}
\nonumber \\ 
&&\times
\prod_{j=1}^{N-7} \frac{E(x^{17}b_4/a_j)E(x^{13}a_j/b_4)} 
{E(x^{3}a_j/b_4)E(x^{27}b_4/a_j)}, \label{bethe4.4}\\
&&\nonumber\\
&&\nonumber\\
\lefteqn{-\omega \left[x^{-10}\frac{E(x^{23}/b_5)}{E(x^{3}/b_5)}\right]^N=
(A_{N-7}B_7)^{5/8}
A_{N-7}^{-2}b_5^{2N-8}(b_1b_2b_4)^{-2}x^{-28-4N}}
\nonumber \\ 
&&\times
\frac{E(b_1/b_5)E(x^{2} b_1/b_5)E(x^4 b_2/b_5)E(x^{2} b_2/b_5)
E(x^4 b_3/b_5)E(x^{6} b_5/b_3)}
{E(x^{8}b_5/b_1)E(x^{10}b_5/b_1)E(x^{8}b_2/b_5)E(x^{6}b_5/b_2)
E(x^{14}b_5/b_3)E(x^{16}b_5/b_3)}
\nonumber \\ &&\times
\frac{E(x^{10} b_4/b_5)E(x^{8} b_4/b_5)E(x^{16}b_6/b_5)E(x^{14} b_6/b_5)
E(x^{19}b_7/b_5)E(x^{17}b_7/b_5)}
{E(x^{2}b_5/b_4)E(b_5/b_4)E(x^{4} b_6/b_5)E(x^{6}b_6/b_5)
E(x^{7}b_7/b_5)E(x^{9}b_7/b_5)}
\nonumber \\ 
&&\times
\prod_{j=1}^{N-7} \frac{E(x^{3}a_j/b_5)E(xa_j/b_5)} 
{E(x^{23}a_j/b_5)E(x^{7}b_5/a_j)}, \label{bethe4.5}\\
&&\nonumber\\
&&\nonumber\\
\lefteqn{-\omega \left[x^{10}\frac{E(x^3 b_6)}{E(x^{23}b_6)}\right]^N=
(A_{N-7}B_7)^{5/8}
A_{N-7}^{-2}b_6^{2N-8}(b_1b_2b_3)^{-2}x^{28+4N}}
\nonumber\\ 
&&\times
\frac{E(x^{6} b_1/b_6)E(x^{8}b_1/b_6)E(x^{8} b_2/b_6)E(x^{10}b_2/b_6)
E(x^{2} b_3/b_6)E(b_3/b_6)}
{E(x^{4}b_6/b_1)E(x^{2}b_6/b_1)
E(x^{2}b_6/b_2)E(b_2/b_6)E(x^{10}b_6/b_3)E(x^8 b_6/b_3)} 
\nonumber \\ &&\times
\frac{E(x^{16} b_4/b_6)E(x^{14}b_4/b_6)E(x^{4} b_6/b_5)E(x^{6}b_6/b_5)
E(x^{7}b_6/b_7)E(x^{23}b_7/b_6)}
{E(x^{4}b_4/b_6)E(x^6 b_4/b_6)E(x^{16}b_6/b_5)E(x^{14} b_6/b_5)
E(x^{13} b_7/b_6)E(x^{17}b_6/b_7)}
\nonumber \\ 
&&\times
\prod_{j=1}^{N-7} \frac{E(x^{23}b_6/a_j)E(x^7 a_j/b_6)} 
{E(x^{3}b_6/a_j)E(x b_6/a_j)}, \label{bethe4.6}\\
&&\nonumber\\
&&\nonumber\\
\lefteqn{-\omega \left[\frac{E(x^6 b_7)}{E(x^{26}b_7)}\right]^N=
(A_{N-7}B_7)^{5/8}
A_{N-7}^{-2}b_7^{2N-10}(b_1b_2)^{-2}}
\nonumber\\ 
&&\times
\frac{E(x^{3} b_1/b_7)E(x^{5}b_1/b_7)E(x^{7} b_2/b_7)E(x^{5}b_2/b_7)
E(x b_7/b_3)E(x^{3}b_7/b_3)}
{E(x^{7}b_7/b_1)E(x^{5}b_7/b_1)E(x^{5}b_7/b_2)E(x^{3}b_7/b_2)
E(x^{13}b_7/b_3)E(x^{11}b_7/b_3)} 
\nonumber \\ &&\times
\frac{E(x^{11} b_4/b_7)E(x^{13}b_4/b_7)E(x^{7} b_7/b_5)E(x^{9}b_7/b_5)
E(x^{13}b_7/b_6)E(x^{17}b_6/b_7)}
{E(xb_4/b_7)E(x^{3}b_4/b_7)E(x^{19}b_7/b_5)E(x^{17}b_7/b_5)
E(x^{7}b_6/b_7)E(x^{23}b_7/b_6)}
\nonumber \\ 
&&\times
\prod_{j=1}^{N-7} \frac{E(x^{26}b_7/a_j)E(x^4 a_j/b_7)} 
{E(x^{6}b_7/a_j)E(x^4 b_7/a_j)}.\label{bethe4.7}
\end{eqnarray}
Taken individually as $x \rightarrow 0$ these equations imply
\begin{eqnarray}
(1-\frac{b_1}{b_5})&=O(x^{N}),& \quad \quad
(1-\frac{b_2}{b_6})=O(x^{N}), \nonumber \\
(1-\frac{b_5}{b_4})&=O(x^{7N}),& \quad \quad
(1-\frac{b_3}{b_6})=O(x^{7N}), \nonumber \\
\frac{(1-\frac{b_5}{b_4})}
{(1-\frac{b_1}{b_5})}&=O(x^{6N}),& \quad \quad
\frac{(1-\frac{b_3}{b_6})}
{(1-\frac{b_2}{b_6})}=O(x^{6N}), 
\end{eqnarray}
which fall into two unconnected groups. Notice also that the seventh equation
(\ref{bethe4.7}) does not provide any link between
$b_7$ and the other $b_i$. We are only able to conclude that
\begin{equation}
b_1=b_4=b_5=\alpha, \quad b_2=b_3=b_6=\beta,\quad \mbox{and} \quad b_7=b,
\end{equation}
with $B_7=\alpha^3\beta^3b$.

\subsection{Mass $m_5$} \label{b5}

We consider the roots
$w_j=a_j$ for $j=1, \ldots, N-6$ and $w_{N-5}=b_1x^{-13}$, 
$w_{N-4}=b_2x^{13}$, $w_{N-3}=b_3x^{-11}$, $w_{N-2}=b_4x^{11}$,
$w_{N-1}=b_5x^{-9}$, $w_N=b_6x^{9}$. In the $x \rightarrow 0$ limit,
the Bethe equations for the latter six roots are 
\begin{eqnarray}
x^{-10N}&=&\frac{-1}{\omega}(A_{N-6}B_6)^{5/8}A_{N-6}^{-2}b_1^{2N-12}
x^{-4N-36}E(b_6/b_1)^{-1}, \nonumber \\
x^{10N}&=&\frac{-1}{\omega}(A_{N-6}B_6)^{5/8}A_{N-6}^{-2}b_2^{2N-12}
x^{4N+36}E(b_2/b_5), \nonumber \\
x^{-10N}&=&\frac{-1}{\omega}(A_{N-6}B_6)^{5/8}A_{N-6}^{-1}b_3^{N-4}b_6^{-2}
x^{-N-52}\left[E(b_3/b_6)E(b_4/b_3)\right]^{-1}, \nonumber \\
x^{10N}&=&\frac{-1}{\omega}(A_{N-6}B_6)^{5/8}A_{N-6}^{-1}b_4^{N-4}b_5^{-2}
x^{N+52}E(b_4/b_3)E(b_5/b_4), \nonumber \\
\left[b_5x^{-9}\right]^{N}&=&\frac{-1}{\omega}(A_{N-6}B_6)^{5/8}
b_5^{4}(b_4b_6)^{-2}
x^{-52}\left[E(b_5/b_4)E(b_2/b_5)\right]^{-1}, \nonumber \\
\left[b_6x^{9}\right]^{N}&=&\frac{-1}{\omega}(A_{N-6}B_6)^{5/8}
b_6^{4}(b_3b_5)^{-2}
x^{52}E(b_6/b_1)E(b_3/b_6).\label{bethe5}
\end{eqnarray}
We conclude from the first, third and sixth of these that
\begin{equation}
b_1=b_6+O(x^{6N}), \quad b_3=b_6+O(x^{3N}), \quad b_3=b_4+O(x^{6N}).
\end{equation}
From the second and fifth we have 
\begin{equation}
b_2=b_5+O(x^{6N}), \quad b_4=b_5+O(x^{3N}. 
\end{equation}
The fourth confirms $b_3=b_4+O(x^{6N})$,
so that we may conclude that the $b_i$ are equal with $B_6=b^6$. 

In terms of the $m_5$ auxiliary functions (\ref{f5}) and (\ref{g5}) for
the product of the six Bethe equations give
\begin{eqnarray}
\lefteqn{\left[\frac{E(x^{19}/b)E(x^{21}/b)E(x^{23}/b)E(x^{29}/b)}
{E(x^{19}b)E(x^{21}b)E(x^{23}b)E(x^{29}b)}\right]^N=}\nonumber \\
&& b^{4N}
\left[\F_5(x^7b)\F^2_5(x^9b)\F^2_5(x^{11}b)\F_5(x^{13}b)\right]^{-1}
\nonumber \\
&& \times \left[\G_5(1/x^7b)\G^2_5(1/x^{9}b)\G^2_5(1/x^{11}b)\G_5(1/x^{13}b)
\right]^{-1}.
\end{eqnarray}
The parameter $b$ is thus found to obey the equation
\begin{eqnarray}
\lefteqn{\left[\frac{E(x^3b)E(x^{9}b)E(x^{11}b)E(x^{13}b)}
{E(x^3/b)E(x^9/b)E(x^{11}/b)E(x^{13}/b)}\right]^N}\nonumber \\
&&\times \left[\frac
{E(x^{33}/b)E(x^{39}/b)E(x^{41}/b)E(x^{43}/b)}
{E(x^{33}b)E(x^{39}b)E(x^{41}b)E(x^{43}b)}\right]^N =b^{4N},
\end{eqnarray}
where the elliptic functions are of nome $x^{60}$.

\subsection{Mass $m_6$} \label{b6}

We consider the roots
$w_j=a_j$ for $j=1, \ldots, N-7$ and 
$w_{N-6}=b_1x^{-14}$, $w_{N-5}=b_2x^{14}$, $w_{N-4}=b_3x^{-10}$,
$w_{N-3}=b_4x^{10}$, $w_{N-2}=b_5x^{-6}$, $w_{N-1}=b_6x^{6}$,
$w_N=b_7x^{16}$. In the $x \rightarrow 0$ limit, the Bethe equations 
for the latter seven roots are
\begin{eqnarray}
x^{-10N}&=&\frac{-1}{\omega}(A_{N-7}B_7)^{5/8}A_{N-7}^{-2}b_1^{2N-12}
b_6^{-2}
x^{-6N-26}E(b_1/b_6)^{-1}, \nonumber \\
x^{10N}&=&\frac{-1}{\omega}(A_{N-7}B_7)^{5/8}A_{N-7}^{-2}b_2^{2N-12}
b_5^{-2}
x^{6N+26}E(b_5/b_2), \nonumber \\
\left[b_3x^{-10}\right]^{N}&=&
\frac{-1}{\omega}(A_{N-7}B_7)^{5/8}b_3^{4}(b_4b_6)^{-2}
x^{-58}E(b_3/b_4)^{-1}, \nonumber \\
\left[b_4x^{10}\right]^{N}&=&
\frac{-1}{\omega}(A_{N-7}B_7)^{5/8}b_4^{4}(b_3b_5)^{-2}
x^{58}E(b_3/b_4),\nonumber \\
\left[b_5x^{-6}\right]^{N}&=&\frac{-1}{\omega}(A_{N-7}B_7)^{5/8}
b_5^{6}(b_2b_4b_6)^{-2}
x^{-40}\frac{E(b_6/b_5)}{E(b_7/b_5)E(b_5/b_2)}, \nonumber \\
\left[b_6x^{6}\right]^{N}&=&\frac{-1}{\omega}(A_{N-7}B_7)^{5/8}
b_6^{6}(b_1b_3b_5)^{-2}
x^{40}\frac{E(b_1/b_6)E(b_6/b_7)}{E(b_6/b_5)}, \nonumber \\
x^{10N}&=&\frac{-1}{\omega}(A_{N-7}B_7)^{5/8}
x^{10N}\frac{E(b_7/b_5)}{E(b_6/b_7)}. \label{bethe6}
\end{eqnarray}
We notice immediately that the third and fourth equations
decouple from the others to give $b_3=b_4+O(x^{10N})$, so
we set $b_3=b_4=\alpha$. From the remaining
five equations we may conclude that (to $O(x^{2N})$ or $O(x^{4N})$)
the remaining $b_i$ are equal, with $B_7=b^5\alpha^2$.

The product of the
seven Bethe equations may be written in terms of the auxiliary
functions (\ref{def6a}) to give
\begin{eqnarray}
\lefteqn{\left[\frac{E(x^{24}/b)E(x^{26}/b)E(x^{20}/\alpha)}
{E(x^{24}b)E(x^{26}b)E(x^{20}\alpha)}\right]^N=}
\nonumber \\
&& b^{4N}
\left[
\frac{F_6(x^{18}b)F_6(x^{24}b)F_6(x^{26}b)F_6(x^{28}b)F_6(x^{30}b)}
{F_6(x^{2}b)F_6(x^{4}b)F_6(x^{6}b)F_6(x^{8}b)F_6(x^{14}b)}
\right]
\nonumber \\
&& \!\times\!
\left[\frac{G_6(1/x^{30}b)G_6(1/x^{28}b)G_6(1/x^{26}b)G_6(1/x^{24}b)
G_6(1/x^{18}b)}
{G_6(1/x^{14}b)G_6(1/x^{8}b)G_6(1/x^{6}b)G_6(1/x^{4}b)G_6(1/x^{2}b)}
\right]
\nonumber \\
&& \!\times \!
\left[\frac{F_6(x^{2}\alpha)F_6(x^{20}\alpha)
F_6(x^{22}\alpha)}{F_6(x^{10}\alpha)F_6(x^{12}\alpha)
F_6(x^{30}\alpha)}
\frac{G_6(1/x^{2}\alpha)
G_6(1/x^{20}\alpha)G_6(1/x^{22}\alpha)}{G_6(1/x^{10}\alpha)
G_6(1/x^{12}\alpha)G_6(1/x^{30}\alpha)}
\right].\nonumber\\
\end{eqnarray}
Now, while some of these factors may be grouped to give the
functions $\F_6$ and $\G_6$ in (\ref{def6b}),
for the others we must use
\begin{eqnarray}
F_6(w)&=&\prod_{l=0}^{\infty}\F_6(x^{12l}w)
\frac{X_{\alpha}(x^{12l+2}w)}{X_{\alpha}(x^{12l+10}w)},
\nonumber \\
G_6(1/w)&=&\prod_{l=0}^{\infty}\G_6(x^{12(l+1)}/w)^{-1}
\frac{Y_{\alpha}(x^{12l+2}/w)}{Y_{\alpha}(x^{12l+10}/w)},
\end{eqnarray}
which is similar to what we did in regime 1.
By applying identities from Appendix A, we find that all factors 
involving $\alpha$ cancel. The final result, 
in terms of elliptic functions of nome $x^{60}$, is
\begin{eqnarray}
\lefteqn{\left[\frac{E(x^6b)E(x^{8}b)E(x^{12}b)E(x^{14}b)}
{E(x^6/b)E(x^{8}/b)E(x^{12}/b)E(x^{14}/b)
}\right]^N}\nonumber \\
&&\times \!\left[\frac{E(x^{36}/b)E(x^{38}/b)E(x^{42}/b)E(x^{44}/b)}
{E(x^{36}b)E(x^{38}b)E(x^{42}b)E(x^{44}b)}\right]^N =b^{4N}.
\end{eqnarray}

\subsection{Mass $m_7$} \label{b7}

For the roots
$w_j=a_j$ for $j=1, \ldots, N-8$ and $w_{N-7}=b_1x^{-14}$, 
$w_{N-6}=b_2x^{14}$, $w_{N-5}=b_3x^{-12}$, $w_{N-4}=b_4x^{12}$,
$w_{N-3}=b_5x^{-10}$, $w_{N-2}=b_6x^{10}$, $w_{N-1}=b_7x^{-8}$,
$w_N=b_8x^{8}$ the last eight Bethe equations in the $x \rightarrow 0$
limit give
\begin{eqnarray}
x^{-10N}&=&\frac{-1}{\omega}(A_{N-8}B_8)^{5/8}A_{N-8}^{-2}
b_1^{2N-16}x^{-6N-32}E(b_8/b_1)^{-1}, \nonumber \\
x^{10N}&=&\frac{-1}{\omega}(A_{N-8}B_8)^{5/8}A_{N-8}^{-2}
b_2^{2N-16}x^{6N+32}E(b_2/b_7), \nonumber \\
x^{-10N}&=&\frac{-1}{\omega}(A_{N-8}B_8)^{5/8}A_{N-8}^{-2}
b_3^{2N-14}b_8^{-2}x^{-2N-62}\left[E(b_3/b_8)E(b_6/b_3)\right]^{-1},
\nonumber \\
x^{10N}&=&\frac{-1}{\omega}(A_{N-8}B_8)^{5/8}A_{N-8}^{-2}
b_4^{2N-14}b_7^{-2}x^{2N+62}E(b_4/b_5)E(b_7/b_4),
\nonumber \\
\left[b_5x^{-10}\right]^{N}&=&\frac{-1}{\omega}
(A_{N-8}B_8)^{5/8}
b_5^{4}(b_6b_8)^{-2}
x^{-72}\left[E(b_5/b_6)E(b_4/b_5)\right]^{-1}, \nonumber \\
\left[b_6x^{10}\right]^{N}&=&\frac{-1}{\omega}
(A_{N-8}B_8)^{5/8}
b_6^{4}(b_5b_7)^{-2}
x^{72}E(b_6/b_3)E(b_5/b_6),\nonumber \\
\left[b_7x^{-8}\right]^{N}&=&\frac{-1}{\omega}
(A_{N-8}B_8)^{5/8}
b_7^{6}(b_4b_6b_8)^{-2}
x^{-62}\left[E(b_7/b_4)E(b_2/b_7)\right]^{-1},\nonumber \\
\left[b_8x^{8}\right]^{N}&=&\frac{-1}{\omega}
(A_{N-8}B_8)^{5/8}
b_8^{6}(b_3b_5b_7)^{-2}
x^{62}E(b_8/b_1)E(b_3/b_8). \label{bethe7}
\end{eqnarray}
In this case we can conclude that the $b_i$ are equal,
to $O(x^{4N})$ or $O(x^{6N})$.

The product of the eight Bethe equations can be written in terms
of the $m_7$ auxiliary functions (\ref{f7}) and (\ref{g7}) as 
\begin{eqnarray}
\lefteqn{\left[\frac{E(x^{18}/b)E(x^{20}/b)E(x^{22}/b)E(x^{24}/b)
E(x^{28}/b)}
{E(x^{18}b)E(x^{20}b)E(x^{22}b)E(x^{24}b)E(x^{28}b)}\right]^N=}\nonumber \\
&& b^{5N}
\left[\F_7(x^6b)\F^2_7(x^8b)\F^2_7(x^{10}b)
\F^2_7(x^{12}b)\F_7(x^{14}b)\right]^{-1}
\nonumber \\
&& \times \left[\G_7(1/x^6b)\G^2_7(1/x^{8}b)\G^2_7(1/x^{10}b)
\G^2_7(1/x^{12}b)\G_7(1/x^{14}b)
\right]^{-1}. 
\end{eqnarray}
Or finally, again with elliptic nome $x^{60}$,
\begin{eqnarray}
\lefteqn{\left[\frac{E(x^4b)E(x^{8}b)E(x^{10}b)E(x^{12}b)E(x^{14}b)}
{E(x^4/b)E(x^8/b)E(x^{10}/b)E(x^{12}/b)E(x^{12}/b)}\right]^N}\nonumber \\
&&\times \left[\frac{E(x^{34}/b)E(x^{38}/b)E(x^{40}/b)E(x^{42}/b)E(x^{44}/b)}
{E(x^{34}b)E(x^{38}b)E(x^{40}b)E(x^{42}b)E(x^{44}b)}\right]^N  = b^{5N}.
\end{eqnarray}

\subsection{Mass $m_8$} \label{b8}

For the roots
$w_j=a_j$ for $j=1, \ldots, N-10$ and $w_{N-9}=b_1x^{-15}$, 
$w_{N-8}=b_2x^{15}$, $w_{N-7}=b_3x^{-13}$,  
$w_{N-6}=b_4x^{13}$, $w_{N-5}=b_5x^{-11}$, $w_{N-4}=b_6x^{11}$,
$w_{N-3}=b_7x^{-9}$, $w_{N-2}=b_8x^{9}$, $w_{N-1}=b_9x^{-7}$,
$w_N=b_{10}x^{7}$
the last ten Bethe equations in the $x \rightarrow 0$
limit give
\begin{eqnarray}
x^{-2N}&=&\frac{-1}{\omega}(A_{N-10}B_{10})^{5/8}A_{N-10}^{-2}
b_1^{2N-20}x^{-20}E(b_{10}/b_1)^{-1}, \nonumber \\
x^{2N}&=&\frac{-1}{\omega}(A_{N-10}B_{10})^{5/8}A_{N-10}^{-2}
b_{2}^{2N-20}x^{20}E(b_{2}/b_9), \nonumber \\
x^{-6N}&=&\frac{-1}{\omega}(A_{N-10}B_{10})^{5/8}A_{N-10}^{-2}
b_3^{2N-18}b_{10}^{-2}x^{-58}\left[E(b_3/b_{10})E(b_8/b_3)\right]^{-1},
\nonumber \\
x^{6N}&=&\frac{-1}{\omega}(A_{N-10}B_{10})^{5/8}A_{N-10}^{-2}
b_4^{2N-18}b_{9}^{-2}x^{58}E(b_4/b_{7})E(b_9/b_4),
\nonumber \\
x^{-9N}&=&\frac{-1}{\omega}(A_{N-10}B_{10})^{\!5/8}\!A_{N-10}^{-1}
b_5^{N-6}(b_8b_{10})^{\!-2}
x^{\!-82}\!\left[E(b_5/b_8)E(b_6/b_5)\right]^{-1}, \nonumber \\
x^{9N}&=&\frac{-1}{\omega}(A_{N-10}B_{10})^{5/8}A_{N-10}^{-1}
b_6^{N-6}(b_7b_{9})^{-2}
x^{82}E(b_6/b_5)E(b_7/b_6),\nonumber \\
\left[b_7x^{-9}\right]^{N}&=&\frac{-1}{\omega}
(A_{N-10}B_{10})^{5/8}
b_7^{6}(b_6b_8b_{10})^{-2}
x^{-82}\left[E(b_7/b_6)E(b_4/b_7)\right]^{-1},\nonumber \\
\left[b_8x^{9}\right]^{N}&=&\frac{-1}{\omega}
(A_{N-10}B_{10})^{5/8}
b_8^{6}(b_5b_7b_{9})^{-2}
x^{82}E(b_8/b_3)E(b_5/b_8),\nonumber \\
\left[b_9x^{-7}\right]^{N}&=&\frac{-1}{\omega}
(A_{N-10}B_{10})^{5/8}
b_9^{8}(b_4b_6b_8b_{10})^{-2}
x^{-68}\left[E(b_9/b_4)E(b_2/b_9)\right]^{-1},\nonumber \\
\left[b_{10}x^{7}\right]^{N}&=&\frac{-1}{\omega}
(A_{N-10}B_{10})^{5/8}
b_{10}^{8}(b_3b_5b_7b_9)^{-2}
x^{68}E(b_{10}/b_1)E(b_3/b_{10}). 
\nonumber \\
\label{bethe8}
\end{eqnarray}
We are able to conclude (to $O(x^{N})$, $O(x^{2N})$, $O(x^{5N})$ or
$O(x^{8N})$) that the $b_i$ are equal, so that $B_{10}=b^{10}$.
In terms of the functions (\ref{f8}) and (\ref{g8}), the product of the
ten Bethe equations gives 
\begin{eqnarray}
\lefteqn{\left[\frac{E(x^{17}/b)E(x^{19}/b)E(x^{21}/b)E(x^{23}/b)E(x^{25}/b)
E(x^{27}/b)}
{E(x^{17}b)E(x^{19}b)E(x^{21}b)E(x^{23}b)E(x^{25}b)E(x^{27}b)}\right]^N=}
\nonumber \\
&& b^{6N}
\left[\F_8(x^5b)\F^2_8(x^7b)\F^2_8(x^9b)\F^2_8(x^{11}b)\F^2_8(x^{13}b)
\F_8(x^{15}b)\right]^{-1}
\nonumber \\
&& \! \times \! \left[\G_8(1/x^5b)\G^2_8(1/x^7b)
\G^2_8(1/x^{9}b)\G^2_8(1/x^{11}b)\G^2_8(1/x^{13}b)\G_8(1/x^{15}b)
\right]^{-1}, \nonumber \\
\end{eqnarray}
which can be written as
\begin{eqnarray}
\lefteqn{\left[\frac{E(x^5b)E(x^{7}b)E(x^{9}b)E(x^{11}b)E(x^{13}b)
E(x^{15}b)}
{E(x^5/b)E(x^7/b)E(x^{9}/b)E(x^{11}/b)E(x^{13}/b)
E(x^{15}/b)}\right]^N}\nonumber \\
&&\times \!\left[\frac{E(x^{35}/b)E(x^{37}/b)E(x^{39}/b)E(x^{41}/b)E(x^{43}/b)
E(x^{45}/b)}
{E(x^{35}b)E(x^{37}b)E(x^{39}b)E(x^{41}b)E(x^{43}b)
E(x^{45}b)}\right]^N \!=\! b^{6N}.
\nonumber \\
\end{eqnarray}
again with elliptic nome $x^{60}$.

\end{document}